\documentclass[12pt,letterpaper]{JHEP3}

\title{Fundamental Superstrings  as Holograms}

\preprint{TIFR/TH/07-15\\IC/2007/039}
\author{Atish Dabholkar\footnote{Email:
atish@theory.tifr.res.in}$^{~1, ~2}$ and Sameer Murthy\footnote{Email:
smurthy@ictp.it}$^{~3}$\\
\it $^1$Department of Theoretical Physics\\
\it Tata Institute of Fundamental Research\\
\it Homi Bhabha Rd, Mumbai 400 005, India\\

\it $^2${Laboratoire de Physique Th\'eorique et Hautes Energies (LPTHE)\\
\it{Tour 24-25, 5\` eme \'etage, Boite 126, 4 Place Jussieu, 75252
Paris Cedex 05}}\\
\it{Unit\'e Mixte de Recherche (UMR 7589)}\\
\it{Universit\'e Pierre et Marie Curie-Paris 6; CNRS; Universit\'e
Denis Diderot-Paris 7}\\

\it $^3$Abdus Salam International Centre for Theoretical Physics\\
\it Strada Costiera 11, Trieste 34014, Italy}

\abstract{ The worldsheet of a macroscopic fundamental superstring in the Green-Schwarz light-cone gauge  is viewed as a possible boundary hologram of the near horizon region of a small black string. For toroidally compactified strings, the hologram has global symmetries of $\bf AdS_3 \times S^{d-1} \times T^{8-d}$ $(d =3, \ldots, 8)$,  only some of which extend to local conformal symmetries.  We construct the bulk string theory in detail for the particular case of $d=3$.  The symmetries of the hologram are    correctly reproduced from this exact worldsheet description in the bulk. 
Moreover, the central charge of the boundary Virasoro algebra obtained from the bulk agrees with the Wald entropy of the associated small black holes. This construction provides an exact CFT description of the near horizon region of small black holes both in Type-II and heterotic string theory arising from multiply wound fundamental superstrings.}

\newcommand{\IR}{\mathbb{R}}
\newcommand{\IZ}{\mathbb{Z}}

\newcommand{\LL}{{\cal L}}
\def\s{\sigma}

\def\a{\alpha}
\def\b{\beta}

\def\h{\eta}

\def\G{\Gamma}

\def\CL{{\cal L}}

\def\CN{{\cal N}}

\def\CP{{\cal P}}

\def\half{{\frac12}}

\def\IC{\relax\hbox{$\inbar\kern-.3em{\rm C}$}}

\def\zb{\bar z}

\def\IC{{\bf C}}

\def\CN{{\cal N}}

\def\bea{\begin{eqnarray}}
\def\eea{\end{eqnarray}}
\def\be{\begin{equation}}
\def\ee{\end{equation}}
\def\ba{\begin{align}}
\def\ea{\end{align}}
\def\bse{\begin{subequations}}
\def\ese{\end{subequations}}
\def\1F1{{}_1\!F_1}
\def\2F0{{}_2\!F_0}

\def\G{\Gamma}
\def\a{\alpha}

\def\h3{$\textrm{H}_3^+$}

\def\IC{{\mathbb C}}
\def\IR{{\mathbb R}}
\def\IZ{{\mathbb Z}}

\def\CP{{\bf CP}}

\newcommand{\beq}{\begin{equation}}
\newcommand{\eeq}{\end{equation}}
\newcommand{\ber}{\begin{eqnarray}}
\newcommand{\eer}{\end{eqnarray}}

\def\be{\begin{eqnarray}}
\def\ee{\end{eqnarray}}

\newcommand{\ta}{{\tilde \alpha}}
\newcommand{\tK}{{\tilde K}}
\newcommand{\tE}{{\tilde E}}

\newcommand{\da}{{\dot a}}

\def\p{\partial}

\def\CP{{\cal P}}

\def\CN{{\cal N}}

\def\CJ{{\cal J}}
\def\CP{{\cal P }}
\def\CL{{\cal L}}
\def\CV{{\cal V}}
\def\CO{{\cal O}}

\def\CK{{\cal K}}
\def\CP{{\cal P }}

\def\CV{{\cal V }}

\def\zb {\bar{z}}

\def\G{\Gamma}

\font\manual=manfnt
\def\dbend{\lower3.5pt\hbox{\manual\char127}}

\def\IZ{\relax\ifmmode\mathchoice
{\hbox{\cmss Z\kern-.4em Z}}{\hbox{\cmss Z\kern-.4em Z}}
{\lower.9pt\hbox{\cmsss Z\kern-.4em Z}} {\lower1.2pt\hbox{\cmsss
Z\kern-.4em Z}}\else{\cmss Z\kern-.4em Z}\fi}
\def\half {{1\over 2}}

\def\p{\partial}
\def\pb{\bar{\partial}}
\def\bar{\overline}

\def\CN{{\cal N}}

\def\rt2{\sqrt{2}}
\def\irt2{{1\over\sqrt{2}}}

\def\s{\sigma}
\def\b{\beta}
\def\a{\alpha}

\font\cmss=cmss10
\font\cmsss=cmss10 at 7pt
\def\IL{\relax{\rm I\kern-.18em L}}
\def\IH{\relax{\rm I\kern-.18em H}}
\def\IR{\relax{\rm I\kern-.18em R}}
\def\inbar{\vrule height1.5ex width.4pt depth0pt}
\def\IC{\relax\hbox{$\inbar\kern-.3em{\rm C}$}}
\def\rlx{\relax\leavevmode}
\def\ZZ{\rlx\leavevmode\ifmmode\mathchoice{\hbox{\cmss Z\kern-.4em Z}}
 {\hbox{\cmss Z\kern-.4em Z}}{\lower.9pt\hbox{\cmsss Z\kern-.36em Z}}
 {\lower1.2pt\hbox{\cmsss Z\kern-.36em Z}}\else{\cmss Z\kern-.4em
 Z}\fi}
\def\IZ{\relax\ifmmode\mathchoice
{\hbox{\cmss Z\kern-.4em Z}}{\hbox{\cmss Z\kern-.4em Z}}
{\lower.9pt\hbox{\cmsss Z\kern-.4em Z}}
{\lower1.2pt\hbox{\cmsss Z\kern-.4em Z}}\else{\cmss Z\kern-.4em
Z}\fi}

\def\zb {\bar{z}}

\def\G{\Gamma}

\def\rt2{\sqrt{2}}
\def\irt2{{1\over\sqrt{2}}}

\def\T{\widetilde}

\def\s{\sigma}

\begin{document}


\section{Introduction}

Consider a `macroscopic' fundamental superstring wrapping $p$ times around a circle of radius $R$ in the limit of large radius. Some spatial directions transverse to the string could be compactified on a torus and the remaining are noncompact. In this case, the worldsheet theory living on such a macroscopic string is particularly simple. For a string winding the circle once, this theory consists of free bosons and free fermions corresponding to the transverse oscillations of the string.  As long as the energy scales of excitations are much smaller compared to the string scale, the macroscopic string cannot break up or emit smaller loops of string. At very weak coupling, these low energy excitations along the string are expected to decouple from the surrounding supergravity fields. Moreover, the free worldsheet theory is manifestly superconformal. These observations raise the question if a fundamental macroscopic superstring could be interpreted as a hologram of some bulk dual theory.

To find the holographic dual, one could examine how the spacetime geometry is modified by the backreaction of the string. The supergravity solution corresponding to such an infinitely extended fundamental superstring was found in \cite{Dabholkar:1989jt,{Dabholkar:1990yf}} using the two-derivative string effective action. A fundamental superstring is in many ways the most basic `solitonic' object in string theory and this solution is the most elementary brane solution in string theory. Indeed, all other p-brane solutions can be constructed from it simply by applying T and S duality transformations to the supergravity fields.

A characteristic property of this solution in all dimensions is that near the core of the string, the effective string coupling $g_s^2$ determined by the local value of the dilaton field goes to zero. This suggests that even after taking into account the backreaction, the worldsheet would continue to decouple from the bulk. On the other hand, the string metric near the core is singular and the curvatures become of the order of the string scale. This suggests that it would be necessary to take into account higher derivative terms in the tree-level string effective action to fully analyze the `geometry' near the core. In fact, since the curvature is of the order of the string scale, corrections arising from various higher derivative terms would be equally important and an exact CFT description would be necessary. One might hope that after taking into account the corrections  to the geometry to all orders in  $\a'$ expansion of the tree level effective action  and possibly exactly by using some bulk worldsheet conformal field theory, it would be possible to obtain the holographic dual of the fundamental string hologram.

Further support for this idea comes from the investigations of higher derivative corrections to the `geometry' and entropy of what have been termed `small' black holes \cite{{Dabholkar:2004yr}, {Dabholkar:2004dq}, {Hubeny:2004ji},  {Sen:2005kj}, {Sen:2005ch}, {Sen:2005pu}, {Dabholkar:2005by}, Dabholkar:2005dt}. If we take the  radius $R$ of the circle along which  the string is wrapping to be very small instead of very large, then one can view the string as a point-like object in one lower dimension. The string can carry in addition some  quantized momentum $q$ along the internal circle. In this case, one obtains a BPS point-like object with two charges $q$ and $p$. From the perturbation analysis of the spectrum one finds that they have exponentially large degeneracy that goes as $\exp{(c \sqrt{pq})}$ as a function of the two charges where the constant $c$ equals $4\pi$ for heterotic strings and $2 \pi \sqrt{2}$ for Type-II strings in all dimensions. It is natural to ask then if there is a two-charged BPS black hole  whose entropy corresponds to the degeneracy of these microscopic states similar to the three-charge case \cite{Strominger:1996sh}.

This expectation is indeed borne out in a number of examples with a beautiful consistency between  the macroscopic and microscopic aspects of the theory. The best studied examples are the heterotic small black holes in four dimensions with $\CN=4$ supersymmetry. These black holes were analyzed using certain F-type four-derivative supersymmetric corrections to the effective action which depend on a particular quadratic contraction of the Riemann tensor. This analysis reveals that upon inclusion of these $\a'$ corrections,  the geometry near the core is no longer singular but is of the form $\bf AdS_2 \times S^1 \times S^2$. The sphere $\bf S^2$ has radius of order one in string units and can be regarded as the `horizon' of this extremal small black hole. The dilaton no longer vanishes at the core and the four-dimensional string coupling $g_4^2 \sim 1/\sqrt{pq}$ is now small but finite. As a result, the area of the horizon measured in units of the four dimensional Planck length is large and scales as $\sqrt{pq}$. The resulting entropy, incorporating the modifications due to Wald \cite{Wald:1993nt,
Iyer:1994ys, Jacobson:1994qe} to the Bekenstein-Hawking formula \cite{Hawking:1971tu,Bekenstein:1973ur}, is  in perfect agreement with the microscopic degeneracy including the precise numerical coefficient.

Inclusion of other higher derivative corrections is expected to correct the geometry further. Moreover, in string theory, the metric, like all other fields, is subject to field redefinitions. Geometric notions at the string scale determined by a given metric are  not invariant under such field redefinitions.  What makes the above analysis tractable and reliable is the fact that the \textit{Wald entropy} of a black hole is a much more robust physical quantity than the `geometry' of the horizon. To begin with, for these black holes, the absolute degeneracy of these states equals a topological index given by a helicity supertrace \cite{Dabholkar:2005by, {Sen:2005pu}, {Sen:2005ch}, {Dabholkar:2005dt}}. Furthermore, the system can be analyzed from a five-dimensional point of view. The radius of the circle $\bf S^1$ of the near horizon region gets attracted to the near horizon value of $\sqrt{q/p}$ in string units irrespective of the asymptotic value $R$ of the radius. The $\bf AdS_2 $ and the $\bf S^1$ factor can then be combined into a fiber bundle as an $\bf AdS_3$ with possible global identifications which could be viewed as the near horizon region of a small black string\footnote{The extent of the string can be infinite and only its thickness is `small' given by the string scale horizon. It is perhaps more accurate to call it a `thin' black string. }.  Using the larger symmetries of $\bf AdS_3$ in this set-up, the Wald entropy can then be related to the anomaly in the boundary R-current and in turn to the bulk Chern-Simons terms \cite{Kraus:2005zm, {Kraus:2005vz}}. These are already included in the four-derivative action and are not further corrected by other higher derivative terms. Thus the Wald entropy computed from the five-dimensional four-derivative supersymmetric action is determined entirely by symmetries and anomalies under the reasonable  assumption that the near horizon region continues to have the symmetries of $\bf AdS_3$ even after including all higher derivative corrections. This reasoning explains why analysis of the four derivative action is adequate for computing certain quantities such as the Wald entropy. One can also show explicitly using the entropy function formalism \cite{Sen:2005kj} that includes higher curvature terms that  Wald entropy is invariant under field redefinitions barring singular ones that take $\bf AdS_3 \times S^2$ to a singular space\footnote{These arguments based on anomalies  also explain why keeping only the four-derivative action is not enough for the small black holes in Type-II string since the gauge current is non-anomalous and is not useful to determine the Wald entropy. We will explain these issues in some detail in $\S{\ref{Near}}$.}.

One can actually go further and compare even subleading corrections to the statistical entropy in an asymptotic expansion in $1/\sqrt{pq}$. The subleading corrections to thermodynamic quantities  are of course  ensemble dependent, but there are finite possibilities to choose from which can be compared with the microscopic counting to determine which is the correct one.  The microscopic counting of these states is exact since it can be done in string perturbation theory. For the macroscopic analysis, one can use the ensemble proposed in \cite{Ooguri:2004zv} or in \cite{Sen:2005pu} with an appropriate measure \cite{Shih:2005he, {Denef:2007vg}}.  One then finds that the macroscopic entropy and  the microscopic entropy are in striking agreement to all orders in an asymptotic expansion which is governed by the same associated Bessel function. Since the asymptotic expansion is  determined entirely by the saddle point quantities, this comparison is independent of subtleties having to do with the choices of contours for inverse Laplace transform that enters the definition of the ensemble. It is nontrivial that the same associated Bessel function appears in the two analyses that are \textit{a priori} completely unrelated. Such a comparison of macroscopic and microscopic entropies  to all orders constitutes a nontrivial check of the consistency of string theory.

Even though the agreement between  the microscopic counting with the Wald entropy is best understood for heterotic small black holes in four dimensions and the corresponding string in five dimensions, there are strong indications that many aspects of the story are true in all dimensions and also for Type-II strings. A general scaling argument due to Sen \cite{{Sen:1995in}} gives the correct dependence $\sqrt{pq}$ of the entropy on the charges in all dimensions  \cite{{Peet:1995pe}} \textit{assuming} that upon inclusion of  the higher derivative corrections the geometry near the core has a black hole horizon. The precise numerical coefficient cannot be computed because the supergravity analysis of higher derivative actions in higher dimensions is more complicated. The important point though is that the scaling argument seems to work uniformly in all dimensions and for all superstrings because it only relies on tree level bosonic action for NS fields that is common to all string theories. The scaling argument can also be successfully generalized to the states with spin
\textit{assuming} that upon inclusion of  the higher derivative corrections the geometry near the core has a black ring horizon. The entropy in this case has the form $\sqrt{pq-rJ}$ with a correct dependence on the spin $J$ and a dipole charge $r$ in agreement with the microscopic counting \cite{Iizuka:2005uv,{Dabholkar:2005qs}, {Dabholkar:2006za}}. 

One could elevate these observations to a general principle that corresponding to every solitonic system in string theory which has a large entropy, there must be a solution realizing a black object  in the low 
energy effective action that has the same entropy. This would include not only big black holes and black rings but also the small ones. The microscopic and macroscopic structure of the theory can then be consistent with each other in a natural way. By the same token, and from the general experience in holography, one expects that any solitonic object with a worldvolume theory which typically will be conformal in deep infrared must have an $\bf AdS$ holographic dual as long as gravity decouples from the worldvolume. This reasoning suggests that corresponding to the worldvolume of the fundamental string,  a holographic dual  must exist in all dimensions.

Encouraged by these  general arguments and some of the successes, we examine in this paper the idea of a fundamental superstring as a hologram,  taking seriously the $\bf AdS_3$ symmetries of the near horizon region. For reasons outlined above, we choose to be guided by symmetries, anomalies, and Wald entropy and refer to the string scale geometry only as a shorthand for signifying the relevant symmetries. We find that these considerations lead us to a very tightly constrained theory describing the worldsheet dynamics of strings in the bulk. This worldsheet theory involves a noncompact WZW model $SL(2)_{k=2}$ (and its heterotic counterpart) which gives us the correct entropy. Precisely for this theory,  we find that the boundary (super)symmetries are realized in the bulk string theory. 

The  discussion is organized as follows. In $\S{\ref{Small}}$ we set up our conventions, review what is known about small black holes, and  list the expected global superconformal symmetries of the near horizon region in this context containing an $\bf AdS_3$ factor. This raises a number of puzzles which we outline and resolve in the subsequent sections. In $\S{\ref{Hologram}}$ we consider the fundamental superstring as a hologram in the Green-Schwarz light-cone formalism. This analysis makes transparent how the global superconformal symmetries could be realized in the hologram and which of them can be extended to local superconformal symmetries. We also discuss an unusual light-cone gauge which is relevant for the comparison with the bulk dual.
In $\S{\ref{Bulk}}$ we specialize to the case of the five-dimensional, Type-II small black string and construct the dual bulk theory with the symmetries of $\bf AdS_3 \times S^2$. In $\S{\ref{Heterotic}}$ we repeat the analysis for five-dimensional heterotic small black strings. In particular, we construct explicitly the boundary symmetries from the bulk, compute the boundary entropy from the bulk, and show that these are in agreement with the hologram. We conclude in $\S{\ref{Conclusions}}$ with a discussion of conclusions, open problems, and outlook.

There a number of related works that have some overlap with considerations here \cite{{Dabholkartobe},{Giveon:2006pr}, {StromingerStrings}, {Johnson:2007du}, {Lapan:2007jx}, {Kraus:2007vu}}. We will comment on some relations to these works during the course of discussion.

\section{Near Horizon Symmetries of Small Black Holes \label{Small}}

\subsection{Macroscopic Superstrings \label{Macro}}

To discuss various toroidally compactified superstrings uniformly, we take the spacetime to be of the form $\IR^{1,1} \times \IR^d \times \textbf{T}^{8-d}$ with coordinates $X^M; M = 0, \ldots, 9$ split as $M = (\mu, i, m)$. The macroscopic string worldsheet extends along the Lorentzian space $\IR^{1,1}$ with coordinates $X^\mu; \mu = 0, 9$ where $X^0$ is the time coordinate and $X^9$ is a circle coordinate, $X^9 \sim X^9 + 2\pi R$. There are $d$ noncompact transverse directions $X^i; i =1, \ldots, d$ along a Euclidean space $\IR^d$ and $(8-d)$ compact directions $X^m; m = d+1, \ldots, 8$ along a torus $\bf T^{8-d}$.

The worldsheet action in conformal gauge for these ten bosonic spacetime coordinates is given by
\begin{equation}\label{bosaction}
    \frac{1}{2\pi \a'} \int d\s d\tau\left( \partial_+ X^M \partial_- X^N \eta_{MN} \right),
\end{equation}
where $\eta_{MN}$ is the 10d Lorentzian metric with signature mostly positive. We have defined $\sigma^{\pm} = \tau \pm \sigma$.
In addition, there are worldsheet fermionic partners appropriate for the heterotic or the Type-II string and leftmoving bosons $H^I$ with $I=1,\ldots, 16$ for the heterotic string that parameterize an internal torus of $E_8 \times E_8$. The total action is subject to Virasoro constraints which we discuss in some detail later in $\S{\ref{Hologram}}$.

Now consider a fundamental string wrapping $p$ times carrying quantized momentum $q$ along the circle. We  define dimensionless left-moving and right-moving momenta
\begin{equation}\label{leftright}
    q_{R, L} = \sqrt{\frac{\a'}{2}}( \frac{q}{R} \pm \frac{p R}{\a'} ).
\end{equation}
If we take the right-movers of the superstring to be in the ground state then this state is supersymmetric and the mass $M$ saturates the BPS bound
\begin{equation}\label{bps}
    M = \sqrt{\frac{2}{\a'}} q_R.
\end{equation}
The left-moving oscillation number $N_L$ of the transverse oscillations satisfies the Virasoro constraint
\begin{equation}\label{nlhet}
    N_L -1 = \frac{1}{2} (q_R^2 - q_L^2) = pq,
\end{equation}
for the heterotic string and
\begin{equation}\label{nlhetII}
    N_L = pq,
\end{equation}
for the Type-II string. There is a large degeneracy $d(q, p)$ of such states since this constraint can be satisfied by exciting various oscillations in many different ways. The statistical entropy given by the logarithm of $d(q, p)$ goes as
\begin{equation}\label{entropy}
    c \sqrt {pq},
\end{equation}
with $c= 4\pi$ for heterotic and $c =2\pi\sqrt{2}$ for Type-II.

In the limit of large $R$ for fixed $q$, this state can be viewed
as an infinitely extended string that will act as the source for various supergravity fields.  Let $r$ be the radial coordinate along the noncompact directions $r^2 = x^i x^i$. The  dilaton field $\Phi(r)$ in the $(d+2)$ noncompact dimensions is given by the transverse harmonic function
\begin{equation}\label{dilaton}
    e^{-2 \Phi(r)} = 1 + \frac{p \Omega }{r^{d-2}}
\end{equation}
where $\Omega$ is a geometric factor. 
The metric in the string frame then takes the form
\begin{equation}\label{metric}
    ds^2 = e^{2\Phi(r)} (dx^\mu dx_\mu) + (dx^i dx^i) + (dx^m dx^m),
\end{equation}
and the nonvanishing components of the  2-form field $B_{MN}$ are given
by
\begin{equation}\label{bfield}
    B_{09} = ( 1 - e^{2 \Phi(r)}).
\end{equation}

\subsection{Small Black Holes, Scaling, and Near Horizon Symmetries \label{Near}}

Taking into account the higher derivative corrections is in general very complicated because one has to solve higher order nonlinear differential equations. The task is greatly simplified using supersymmetry. In four dimensions using the superconformal formulation of higher derivative supergravity one can incorporate four-derivative F-type terms and find the BPS solutions \cite{LopesCardoso:1998wt,LopesCardoso:1999cv,LopesCardoso:1999xn, LopesCardoso:1999ur, LopesCardoso:2000qm}.
The solutions corresponding to two-charge heterotic BPS states discussed above  are found to have a string scale near horizon geometry of $\bf AdS_2 \times S^1 \times S^2$ \cite{Dabholkar:2004yr, Dabholkar:2004dq}. This system can also be analyzed from a five-dimensional point of view using the four derivative supergravity action of \cite{Hanaki:2006pj}.
The corresponding small black string solution with an $\bf AdS^3 \times S^2$ near horizon geometry is discussed in  \cite{Castro:2007sd, Castro:2007hc}.

The main virtue of the four derivative action in five dimensions is that it already incorporates the gravitational Chern-Simons interaction and all terms related to it by supersymmetry. Under suitable conditions which will be discussed in greater detail in $\S{\ref{Wald}}$, one can determine the Wald entropy of the black holes completely using symmetries and anomalies. The four derivative action is thus adequate to be able to draw reliable and useful  conclusions about the entropy of the heterotic small black holes.

The attractor values of the dilaton and the radius are determined entirely in terms of charges
\begin{equation}\label{attr}
    g_5^2 = \frac{1}{p}, \quad g_4^2 = \frac{1}{\sqrt{pq}}, \quad R = \sqrt{\frac{q}{p}},
\end{equation}
where $g_5$ is the 5d string coupling, $g_4$ is the 4d string coupling, and $R$ is the radius of the circle in string units around which the string wraps. This shows in particular, that for large $p$, the near horizon string coupling can be made arbitrarily small. One can therefore consistently assume that the worldsheet of the fundamental string, which we will later interpret as the hologram, decouples from the massless supergravity fields.

Let us now list the symmetries of this near horizon solution for the heterotic small black string. To start with, we expect to have the global symmetries of $\bf AdS_3 \times S^2$ which are $SL(2, \IR) \times SL(2, \IR)  \times SO(3)$. We also expect a local conformal symmetry $Virasoro \times Virasoro$ from a Brown-Henneaux construction \cite{Brown:1986nw}. The string is a half-BPS state to start with so we have eight unbroken global spacetime supersymmetries. Near the horizon, in the $\CN =2$ formalism that we have used, the supersymmetry is enhanced to include $4$ additional superconformal symmetries. So we expect altogether at least $12$ superconformal symmetries and possibly $16$ superconformal symmetries if the problem could be analyzed in a manifestly $\CN =4$ formalism. In the Type-II case, if a small black hole were to exist, we would expect at least $12 + 12 $ and possibly $16 + 16$ superconformal symmetries. 

As mentioned in the introduction, a general scaling argument suggests that a small black hole ought to exist in all dimensions \cite{Sen:1995in, {Peet:1995pe}}.
If a small black string were to exist in higher dimensions for $\IR^{1,1}\times \IR^d \times \textbf{T}^{8-d}$ compactifications with $d=3, \ldots 8$, we would expect  possible near horizon geometries that have symmetries of $\bf AdS^3 \times S^{d-1} \times T^{8-d}$ with $d=3, \ldots, 8$\footnote{The case $d=2$, although not considered here, could be relevant for studying  small black rings in four dimensions if they exist.}. If we assume that there is a left-moving Virasoro and a right-moving Virasoro as it happens for the D1-D5 system, then we expect to have for the right-movers at least a global $SL(2, \IR)$ symmetry.
The supercharges must transform under $Spin(d) \times Spin(8-d)$ and so we are led to look for a supergroup that contains the bosonic symmetry
$SL(2, \IR) \times Spin(d) \times Spin(8-d)$ and at least $12$ and possibly $16$ global superconformal supersymmetries\footnote{We would like to thank A. Strominger for valuable discussions pertaining to symmetries. See also \cite{StromingerStrings}.}. Possible supergroups that contain sixteen supersymmetries are limited in number. The list of symmetries of heterotic small black strings with a possible supergroup containing them is summarized in the table (\ref{Global}).
\begin{table}
  \centering
  \begin{tabular}{|c|c|c|c|}
  \hline
\textbf{Space}&  \textbf{Horizon Symmetry} & \textbf{R Symmetry} & \textbf{Possible Supergroup}    \\
\hline
$\IR^{8}$ & $Spin(8)$   & $Spin(8)$ & $OSp(8|2)$ \\
\hline
$\IR^{7} \times S^1$   & $Spin(7)$  & $Spin(7)$ & $F(4)$ \\
\hline
$\IR^{6} \times T^2$  & $Spin(6)$  &  $Spin(6) \times Spin(2)$ & $SU(1,1|4)$\\
\hline
$\IR^{5} \times T^3$ & $Spin(5)$ &  $Spin(5) \times Spin(3)$ & $OSp(4^{*}|4)$\\
\hline
$\IR^{4} \times T^4$  & $Spin(4)$ &  $Spin(4) \times Spin(4)$ & $SU(1,1|2)^2$ \\
\hline
$\IR^{3} \times T^5$  & $Spin(3)$  &  $Spin(3) \times Spin(5)$ & $OSp(4^{*}|4)$\\
\hline
\end{tabular}
\caption{For heterotic compactifications of the form $\IR^{1,1}\times \IR^d \times \textbf{T}^{8-d} (d=3, \ldots 8)$, the expected global rotational symmetry of the horizon  in various dimensions are listed here in the second column, and the global R-symmetries of the supercharges are in the third column. Possible supergroups that contain these symmetries assuming $16$ superconformal symmetries are listed in the fourth column. }\label{Global}
\end{table}
For example, the group $OSp(8|2)$ contains $Spin(8) \times Sp(2)$ as a bosonic subgroup and the fermionic generators transform as vector of $Spin(8)$ and a doublet of $Sp(2) \sim SL(2)$. Similarly $OSp(4^{*}|4)$ contains $SO^*(2, 2) \sim SL(2) \times Spin(3)$ and $Sp(4) \sim Spin(5)$ as bosonic subgroups. See \cite{Britto-Pacumio:1999ax} for a nice introduction to supergroups in this string theory context.

There are a number of puzzles that arise from these identifications of the supergroups for the global symmetries of the horizon. It is well-known that maximal allowed local superconformal symmetries are  given by an $\CN = (4, 4)$ superconformal theory that has $SU(2)$ R-symmetry both on the left and on the right. This algebra of local currents has a closed subalgebra whose fermionic part consists of $8+8 =16$ global  superconformal  charges\footnote{Following standard convention, $\CN =(0, 4)$ supersymmetry in two dimensions has four usual right-moving supersymmetries. Together with the four additional superconformal symmetries, it has a total of eight global supercharges.}. How does one square this the much larger global symmetries which for example require $16 +16 =32$ supersymmetries for the Type-II case? For these reasons, we will not commit ourselves to the supergroups in table (\ref{Global}) and regard them as a tentative identification. We will be guided instead by the holograms discussed in $\S{\ref{Hologram}}$ where it is easy to write down the symmetry algebras quite explicitly. The question of global and local symmetries is somewhat subtle even in the hologram and we shall discuss this issue in more detail in $\S{\ref{Hologram}}$. The usual global supersymmetries are easy to display but the realization  of global superconformal symmetries involves an analog of spectral flow. It is not possible to make all global and local symmetries manifest at the same time.

\subsection{Wald Entropy and Anomalies\label{Wald}}

We now briefly review the arguments that utilize the $\bf AdS_3$ symmetry and anomalies to compute the Wald entropy \cite{Kraus:2005zm,Kraus:2005vz,Kraus:2006wn}.

The dynamics of the theory in this background will be governed by an effective three dimensional action, obtained by compactifying the remaining directions including the angular coordinates of the horizon. This effective action will have the form
\begin{equation}
\label{ecomp6} \int d^3 x \,
\sqrt{-\det G} \, (\LL_0^{(3)} + \LL_1^{(3)})\, ,
\end{equation}
where
$\LL_0^{(3)}$ is a lagrangian density with manifest general
coordinate invariance, and $\sqrt{-\det G}\, \LL_1^{(3)}$ denotes
the gravitational Chern-Simons term: \be \label{ecomp7} \sqrt{-\det
G}\, \LL_1^{(3)} = K\, \Omega_3\, , \ee $\Omega_3$ being the Lorentz
Chern-Simons 3-form and $K$ is a constant. The action admits an $\bf AdS_3$ solution
 \begin{equation}\label{ecomp3}
ds_3^2 = L^2 \left[ e^2 (- r^2 dt^2 + r^{-2} dr^2) +
(dy + erdt)^2 \right]\, ,
\end{equation}
where $L$ is an overall scale. We have written it in the form of a fiber bundle. The fiber is a circle with coordinate $y$ and the base is an $\bf AdS_2$ with coordinates $(r, t)$ so that $e$ can be viewed as a unit of charge associated with the Kaluza-Klein reduction along $y$.
One can then show, both
in the Euclidean action formalism
\cite{Kraus:2005zm,Kraus:2005vz,Solodukhin:2005ah} as well as using
Wald's formula \cite{Saida:1999ec,Sahoo:2006vz}, that the entropy of
the black hole with near horizon geometry described in
(\ref{ecomp3}) has  the form: \bea \label{ecomp8} S_{BH} &&= 2\pi
\sqrt{ c_L \, Q\over 6} \quad \hbox{for} \, Q>0\, ,
\nonumber \\
&&= 2\pi \sqrt{c_R \, |Q|\over 6} \quad \hbox{for} \, Q <0\, , \eea
where $Q$ is the electric charge associated with the Kaluza-Klein gauge field, and \be \label{ecomp9} c_L = 24\, (-g(l)+\pi\, K)\, ,
\qquad c_R = 24\, (-g(l)-\pi\, K) \, , \ee \be\label{ecomp10}
g(l) = {1\over 4}\, \pi \, l^3 \, \LL_0^{(3)}, \qquad l =
2Le \, . \ee $\LL_0^{(3)}$ in (\ref{ecomp10}) has to be
evaluated on the near horizon background (\ref{ecomp3}). This gives
a concrete form of the $Q$ dependence of the entropy in terms of the
constants $c_L$ and $c_R$.

The constants $c_L$ and $c_R$ given in (\ref{ecomp9})
can be interpreted
as the left- and right-moving central charges of the two dimensional
CFT living on the boundary of the
$\bf AdS_3$ \cite{Kraus:2005vz,Kraus:2005zm,Solodukhin:2005ah}.
The Kaluza-Klein momentum $Q$ is interpreted as the momentum in this boundary CFT which is the  $(\CL_0 - \bar \CL_0$) eigenvalue
of a given state in this CFT. The two cases in (\ref{ecomp8}) correspond to ${\bar \CL_0} =0$ and $Q > 0$ or $\CL_0 =0$ and $Q< 0$.
With these identifications, (\ref{ecomp8}) can be interpreted as simply Cardy formula in this
CFT. This argument can be summarized by saying that Wald entropy of the bulk equals the Cardy entropy of the boundary.

If the theory has at least $\CN = (0, 2)$ supersymmetry then one can actually do more and determine even $c_L$ and $c_R$ using anomalies.
In our case the boundary theory will in fact  have $\CN = (0,4)$ supersymmetry. In this case, the central charge $c_R$ is related to the central charge of an $SU(2)_R$ current algebra which is also a part of the $\CN = (0,4)$ supersymmetry algebra. Associated with the $SU(2)_R$ currents there will be $SU(2)$ gauge fields in the bulk, and the central charge of the $SU(2)_R$ current algebra will be determined in terms of the
coefficient of the gauge Chern-Simons term in the bulk theory. This
determines $c_R$ in terms of the coefficient of the gauge Chern-Simons
term in the bulk theory \cite{Kraus:2005vz,Kraus:2005zm, {Kraus:2006wn}}.
On the other hand from (\ref{ecomp9}) we see that
$c_L-c_R$ is determined in terms of the coefficient $K$ of the
gravitational Chern-Simons term.
Since both $c_L$ and $c_R$ are determined in terms of the
coefficients of the Chern-Simons terms in the bulk theory, they do
not receive any higher derivative corrections. This completely
determines the entropy from (\ref{ecomp8}). Furthermore the
expression for the entropy derived this way is independent of all
the near horizon parameters and hence also of the asymptotic values
of all the scalar fields. Since this argument is quite general and three-dimensional, it is expected to work for higher dimensional small black strings as well with transverse space of the type $\bf S^{d-1} \times T^{8-d}$ as long as the $Spin(d)$ symmetry couples chirally to bulk gauge fields \cite{Krausunpub}.

While the argument works beautifully for heterotic small black strings, it appears to fail spectacularly for Type-II small black strings. For instance, for Type-II on $\bf T^6$, the F-type four-derivative terms are zero and hence to this order the horizon continues to be singular with a vanishing horizon and the resulting Wald entropy would appear to be zero. What is worse, if we identify isometries of $\bf S^{d-1}$ with the \textit{conformal} $R$-symmetries then one would conclude by  similar reasoning that $c_L = c_R =0$ giving vanishing entropy in contradiction with microscopic counting and the scaling argument.

A correct interpretation of these results, as we will argue in the next two sections is that in the Type-II case, the geometric rotational symmetry of the horizon is nonchiral and does not correspond to the conformal R-symmetry.  There are additional chiral gauge symmetries of stringy origin which can be identified with the conformal R-symmetries both for the right and the left-movers.  The geometric symmetries of the horizon are a nonchiral linear combination of these R-symmetries. This structure will be quite clear also from the hologram that we discuss in
$\S{\ref{Hologram}}$. Now, the coefficient of the gravitational Chern-Simons  term is proportional to $c_L - c_R$ which vanishes. The gauge Chern-Simons term in supergravity being nonchiral also couples to $c_L - c_R$.  Therefore, unlike the heterotic case, the Chern-Simons terms are not useful for determining the entropy\footnote{We would like to thank P. Kraus for explaining this point of view.}.  This means in particular that analyzing only the four-derivative action is not adequate to find the correct entropy but one must take into account all $\alpha'$ corrections as suggested by the scaling argument. Application of the scaling argument will then tell us that entropy will have the right dependence on charges but determination of the precise coefficient is intrinsically stringy and not easily doable in supergravity. This explains why small black holes and black strings have been difficult to find in the Type-II case. Our stringy construction $\S{\ref{Bulk}}$  will give a way to compute this entropy using an exact CFT construction of the worldsheet.

Many of these confusing issues are neatly resolved by looking at the holograms that we expect for this system. We therefore turn next to the hologram for some guidance about the structure of various symmetries.

\section{The Fundamental Superstring as a Hologram \label{Hologram}}

There is a simple way to realize all the required symmetries expected for the near horizon of a small black string using a free field representation which is furnished by the worldsheet of a toroidally compactified Green-Schwarz \textit{macroscopic} superstring in a particular light-cone like gauge. For our purposes, this  specific free field representation is not only simple but will have a direct physical interpretation as the boundary hologram. It is an instructive exercise to work out this representation in some detail. In particular, it will illuminate  the role of global and local symmetries and will provide some guidance as to which of the global symmetries can become local conformal symmetries. We would like to regard all transverse oscillations as the fields along the worldsheet and also solve the Virasoro constraints. For this purpose it will be useful to choose a slight variant of the usual light cone gauge using the compact $X^9$ direction as one of the light-cone coordinates. We discuss this `compact light-cone gauge' and the various resulting algebras in $\S{\ref{Compact}}$. In $\S{\ref{Internal}}$ we choose a further variation of this gauge by using one of the internal compact directions to be the light-cone directions. This will prove useful for later comparison with the bulk holographic dual.

A theory of $p$ identical strings has also a symmetry $S^p$ which permutes the different strings. The full holographic theory is then a symmetric product of $p$ strings. This is consistent with $S$ duality which maps it to a theory of $D$-strings. In the bulk theory which we discuss in the following sections, the corresponding statement is that there are states with non-zero values of spectral flow number in $SL(2)$. In this section, we shall discuss the symmetries of the system for which it is sufficient to consider the free theory of the transverse oscillations of the string.

\subsection{Holograms in the Compact Light Cone Gauge \label{Compact}}

The action of the superstring in the conformal gauge is subject to Virasoro constraints
\begin{eqnarray}
  T_{++} &=& \frac{1}{\a'}\partial_+X^M \partial_+ X_M + T^{int}_{++}=0, \\
     T_{--} &=& \frac{1}{\a'} \partial_-X^M  \partial_-X_M  +  T^{int}_{--}=0.
\end{eqnarray}
Here $T^{int}_{++}$ and $ T^{int}_{--}$ are the stress tensor components of the fermionic and internal coordinates which we discuss more explicitly later.

We would like to solve these constraints explicitly so that we have to deal with only the transverse physical oscillations. For this purpose, it is useful to  define the `compact' light-cone coordinates
\begin{equation}\label{lccoord}
    X^{\pm} = \frac{( X^0 \pm X^9)}{\sqrt{2}}.
\end{equation}
Ignoring the oscillators, the zero mode expansion of the two fields
$X^0$ and $X^9$ is given by
\begin{eqnarray}
  X^0 &=& x^0 + p^0 \a'  \tau \\
  X^9 &=& x^9 +  \a' \frac{q}{R}  \tau  + \frac{w R}{\a'} \sigma
\end{eqnarray}
We now choose the following light-cone like gauge
\begin{equation}\label{clc}
    X^{+} = x^+ + \frac{\a'}{\sqrt{2}}\left[( p^0+ \frac{q}{R})  \tau  + \frac{w R}{\a'} \sigma\right] \, 
\end{equation}
so that the $X^+$ coordinate has no oscillators.
Note that this is different from a discrete light cone. Since the coordinate $X^9$ is compact, $X^9 \sim X^9 + 2\pi R$, the light-cone direction spirals around the cylindrical $(X^0, X^9)$ space but has infinite extent.
As in (\ref{leftright}), we can define dimensionless left-moving and right-moving  lightcone momenta,
\begin{equation}\label{leftrightLC}
    q^{+}_{R, L} = \sqrt{\frac{\a'}{2}}( p^0 + \frac{q}{R} \pm \frac{w R}{\a'} ),
\end{equation}
so that
\begin{equation}\label{xplus}
    \partial_+ X^+ = \sqrt{\frac{\a'}{2}} q^+_R, \qquad \partial_- X^+ = \sqrt{\frac{\a'}{2}}q^+_L.
\end{equation}
This allows us to solve the Virasoro constraints for the remaining longitudinal mode in terms of the transverse modes,
\begin{equation}\label{xminus}
    \partial_+X^- = \frac{\sqrt{\a'}} {q^+_L} ( T^{tr}_{++}); \quad \partial_-X^- = \frac{\sqrt{\a'}} {q^+_R} ( T^{tr}_{--}),
\end{equation}
where the superscript $tr$ refers to all spacetime and internal degrees of freedom that are transverse to $(X^0, X^9)$.
The mass-shell conditions (\ref{bps}), (\ref{nlhet}), and (\ref{nlhetII}) follow from identifying the Fourier modes $\alpha^-_0$ with ${q^-_R}$ and ${\tilde \alpha}^-_0$ with ${q^-_L}$.
We see that  in the limit of $R \rightarrow \infty$, in the zero winding sector $p=0$, this gauge reduces to the usual light-cone gauge. On the other hand, for nonzero winding $p \neq 0$ and for fixed $q$, it resembles the static gauge which is what we are interested in. Unlike the static gauge however, it has the virtue of the light-cone gauge that the Virasoro constraints are explicitly solvable\footnote{A gauge-fixing of this kind was discussed in \cite{Kanitscheider:2006zf}. We thank Kostas Skenderis for bringing this to our attention.}. For the naive static gauge $X^9 = \sigma R$, the Virasoro constraints are quadratic in the $X^0$ oscillators and are difficult to solve at the level of quantum operators.

Since all longitudinal degrees of freedom are now either gauge fixed or determined in terms of the transverse modes, we can focus on the physical transverse modes. Fermions can be incorporated in the usual way and we will use the light-cone Green-Schwarz formalism.  The transverse  action for the Type-IIB superstring on $\IR^{1,1} \times \IR^d \times \textbf{T}^{8-d}$ compactification  is given by
\begin{equation}\label{type2action}
    S = \frac{1}{2\pi \a'} \int d\sigma d \tau (\partial_+X^i \partial_-X^i + \partial_+X^m \partial_-X^m + i\a' S_+^{a \alpha} \partial_-S_+^{a \alpha}  + i\a' \tilde{S}_-^{a \alpha} \partial_+{\tilde S}_-^{a \alpha}),
\end{equation}
where $i = 1, \ldots, d$ transforms in the vector representation of $Spin(d)$; $m = d+1, \ldots, 8$ transforms in the vector representation of $Spin(8-d)$. Similarly, for fermions, $a$ transforms as a spinor of $Spin(d)$ and $\alpha$ as a spinor of $Spin(8-d)$. For uncompactified string, that is, the special case when $d=8$, bosons transform in the  vector representation $\bf 8_v$ and the fermions in the spinor representation $\bf 8_s$ of $Spin(8)$. For the $Spin(8)$ spinor and gamma matrices we follow the conventions in Appendix 5.B of \cite{Green:1987sp}. 

The group-theoretic structure of bosons and fermions for $d<8$ is then determined by the embedding $Spin(d) \times Spin(8-d) \subset Spin(8)$. 
For example, for the $Spin(3) \times Spin(5)$ case, our conventions are summarized in Appendix \ref{B}. For Type-IIA strings, the action is similar except that the left-moving fermions transform in conjugate spinor representation $\bf 8_c$ of $Spin(8)$. For heterotic strings, we do not have left-moving fermions but instead  left-moving internal bosons $H^I; I = 1, \ldots 16$ that live on an $E_8 \times E_8$ lattice.

\subsection{Symmetries of the Holograms \label{symholo}}

We would now like to view this theory defined by (\ref{type2action}) as a hologram and in particular understand all its symmetries. Various global and local symmetries are very easy to work out because the computations are identical to those that appear in the first quantization of the light-cone superstring. The physical interpretation of these symmetries here  is however completely different. The theory given by the action above should be viewed as the `second quantized' string field theory action of the strings moving in the holographically dual bulk theory. Construction of the holographically dual worldsheet which we discuss in $\S{\ref{Bulk}}$ gives  the first quantized realization of this symmetry algebra.

We consider here the special case $d=8$ to simplify the discussion and also  focus on only right-movers while discussing the chiral currents.
The mode expansions of the basic fields are
\begin{equation}\label{xexp}
    \partial_+X^i = \sqrt{\frac{\a'}{2}} \sum_{n \neq 0} \a_n e^{-in\s^+}, \quad 
    \partial_-X^i = \sqrt{\frac{\a'}{2}} \sum_{n \neq 0} \ta_n e^{-in\s^-},
\end{equation}
for bosons and
\begin{equation}\label{sexp}
    S^a_+ = \sum_{n=-\infty}^\infty S_n^a e^{-in \s^+}, \quad
    S^a_- = \sum_{n=-\infty}^\infty {\tilde S}_n^a e^{-in \s^-},
\end{equation}
for the fermions. The oscillators satisfy the usual canonical commutation relations
\begin{equation}\label{ccr}
    [\a^i_m, \a^j_n] = m \delta_{m+n} \delta^{ij}, \quad \{S^a_m, S^b_n\} = \delta_{m+n} \delta^{ab},
\end{equation}
and similarly for the left-movers. In addition there are bosonic zero modes $x^i$ and $p^i$ which satisfy the Heisenberg commutation relations
\begin{equation}\label{heisen}
    [x^i, p^j] = i \delta^{ij}.
\end{equation}

To begin with, the action has a global $Spin(8)$ rotational symmetries generated by $J^{ij}$, which we write as
\begin{equation}\label{J}
    J^{ij} = L^{ij} + E^{ij} + K_0^{ij} + \tE^{ij} + \tK_0^{ij},
\end{equation}
with
\begin{equation}\label{L}
    L^{ij} = (x^i p^j - x^j p^i),
\end{equation}
\begin{equation}\label{E}
    E^{ij} = -i \sum_{n >0} \frac{1}{n} (\a_{-n}^i \a_n^j -\a_{-n}^j \a_n^i),
\end{equation}
\begin{equation}\label{K0}
    K_0^{ij} = -\frac{i}{4} \sum_{n>0} S^a_{-n} \gamma^{ij}_{ab} S^b_n,
\end{equation}
and similarly for the contributions $\tE^{ij}$ and $\tK_0^{ij}$ from the left-moving oscillators. Note that even though the oscillator contributions are chiral,  the piece $L_{ij}$ which depends on the zero modes $x^i$ and $p^i$ is nonchiral and as a result the rotation symmetry generated by the $J_{ij}$ is nonchiral. This fact will be important later.

In addition to this global, nonchiral symmetry, there are a large number of local, chiral symmetries.  For the right-movers, we have the conformal symmetries  generated by
the spin-2 stress tensor $T(\sigma^+)$, supersymmetries generated by the spin-3/2 currents $Q^{\dot a}(\s^+)$  as well as  $Spin(8)$ affine algebra generated by the spin-1 currents $K^{ij}(\sigma^+)$.   These operators are given by
\begin{equation}\label{stresstensot}
    T(\sigma^+) = \frac{1}{\a'}( \partial_+X^i \partial_+X^i )+
    \frac{i}{2} S^a_+ \partial_+ S^a_+
\end{equation}
\begin{equation}\label{super}
    Q^{\da} (\sigma^+) = \frac{1}{\sqrt{\a'}} (\gamma^i)^{\da}_{a} S_+^a \partial_+X^i,
\end{equation}
\begin{equation}\label{spin}
    K^{ij}(\s^+) = - \frac{i}{4} S^a_+ \gamma^{ij}_{ab}S^b_+ \, .
\end{equation}
The index $i$ transforms in the vector representation $\bf 8_v$ of $Spin(8)$, the index $a$ in the Majorna-Weyl spinor representation $\bf 8_s$ of positive chirality, and the index $\dot a$  in the conjugate Majorana-Weyl spinor representation $\bf 8_c$ of negative chirality and $\gamma^i_{a \da}$ are the Clebsch-Gordon coefficients between these three representations. There are similar currents for the left-movers. In the heterotic case, one does not have the supersymmetries and the $Spin(8)$ current algebra  on the left but instead the $E_8 \times E_8$ current algebra which contributes also to the stress tensor as usual.

Using the mode expansions of operators above and the commutation relations (\ref{ccr}), it is easy to obtain the Virasoro algebra
\begin{eqnarray} \label{Vira}
[\CL_{m}, \CL_{n}] = (m-n)\CL_{m+n} + {c \over 12} (m^{3}-m) \delta_{m+n},
\end{eqnarray}
with central charge $c =12p$\footnote{For a singly would string the central charge would be $12$ but multiple winding is equivalent to having $p$ species and hence the central charge becomes $12p$.}, the commutators
\begin{eqnarray} \label{LG}
[\CL_m , Q_n^{\dot a} ] &=& (\frac{1}{2} m -n) Q_{m+n}^{\dot a} \, , \quad
[\CL_m , K_n^{ij}] = -n  K_{m+n}^{ij},
\end{eqnarray}
and  the Kac-Moody algebra
\begin{equation}\label{LK}
    [K^{ij}_m , K_n^{kl}] = i \delta^{ik} K_{m+n}^{jl}
    - i \delta^{il} K_{m+n}^{jk} + \frac{\tilde k}{2}(m -n)  \delta_{m+n} (\delta^{ik}\delta^{jl} - \delta^{jk}\delta^{il}).
\end{equation}
There is in addition a nontrivial anticommutator \cite{Green:1980zg},
\begin{equation}\label{GG}
    \{ Q^\da_m , Q^{\dot b}_n \} = 2 \delta^{{\dot a} {\dot b}} \CL_{m+n} + (m -n) (\gamma^{ij})^{{\dot a}{\dot b}}  K^{ij}_{m+n} + {\tilde c} (m^2 - \frac{1}{4})  \delta^{{\dot a} {\dot b}} \delta_{m+n},
\end{equation}
where $\tilde k$ and $\tilde c$ are some constants.
The modes of the supercurrent $Q^\da_m$ transforms as a spinor under the global rotations,
\begin{equation}\label{JQ}
   [ J^{ij},  Q^\da_n ] = - \frac{i}{2} (\gamma^{ij})^{\da}_{{\dot b}} Q^{\dot b}_n.
\end{equation}
If we consider a toroidal compactification $\IR^{1,1} \times \IR^d \times \textbf{T}^{8-d}$, we have noncompact coordinates $x^i; i = 1, \ldots, d$ and compact coordinates $x^m; m = d+1, \ldots, 8$. The toroidal identifications of the compact coordinates $x^m$ breaks the rotational symmetry and we only have the $Spin(d)$ symmetry generated by $J_{ij}$ now with $i, j = 1, \ldots d$. However, the chiral symmetries above work as before simply by decomposing the spinor indices of $Spin(8)$ under $Spin(d) \times Spin(8-d)$.

There is an anomaly free subalgebra of the Virasoro algebra generated by
$(\CL_0, \CL_{1}, \CL_{-1})$  and $({\tilde \CL}_0, {\tilde \CL}_{1}, {\tilde \CL}_{-1})$ which generates the global $SL(2, \IR) \times SL(2, \IR)$ which can be identified with the isometries of an $\bf AdS_3$. We also have the rotational $Spin(d)$ symmetries generated by $J_{ij}$ which can be identified with the isometries of a spherical `horizon' $S^{d-1}$. There are sixteen supersymmetries $(Q^\da_0, {\tilde Q}^\da_0)$. In addition, there can be conformal supersymmetries which we will discuss shortly. These symmetries together already give us enough reason to consider this worldsheet theory as the hologram of the near horizon geometry of a small black string in $d+1$ noncompact dimensions discussed in the previous subsection that we are after.
Taking this hologram seriously then predicts that the bulk theory must have not just these global symmetries but also the local Virasoro symmetries on the left and the right as well as additional chiral symmetries which we now discuss.

The hologram also makes it transparent as to which symmetries can possibly be realized as chiral, local symmetries and which are only global symmetries. For example, it is clear that even though this algebra looks very close to a possible superconformal algebra of $(8, 0)$ type with a possible $Spin(8)$ conformal R-current, this is not true. This is because the commutator of $K^{ij}_m$ with $Q^\da_n$ does not close and one obtains instead,
\begin{equation}\label{KQ}
    [K^{ij}_m, Q^\da_n ] = -\frac{i}{4} (\gamma^{ij} \gamma^k)_{a\da} \sum_{r} S^a_{-r} \alpha_{n+m+r}^k ,
\end{equation}
so we see that the right-hand side does not equal $\frac{i}{2} \gamma^{ij}_{\da {\dot b}} Q^{\dot b}_{m+n}$ as one might expect if this were to form a closed algebra and if $Q^\da_m$ were to commute as modes of a spinor operator under the R-symmetry.
The reason for this failure is of course obvious since the generators $Q^\da_m$ commute as spinors only with  the total angular momentum $J_{ij}$ and not if we consider only $K_0^{ij}$. More explicitly, $Q^\da_m$ defined in (\ref{super})  contain terms that are proportional to $p^i$ which commute with $K_0^{ij}$ and commute as   a vector only when we take into account the orbital angular momentum $L_{ij}$.

This shows that even though we have a global $Spin(8)$ R-symmetry that acts on the supercharges, it cannot be extended to a local, chiral conformal R-symmetry \cite{Seiberg:1997ax}. This is just as well because otherwise one would obtain a closed $\CN =(0, 8)$ \textit{superconformal}  algebra from the commutators  of $(\CL_m, Q^\da_n, K^{ij}_l )$ with $Spin(8)$ chiral R-symmetry. This would contradict general theorems which state that the maximal allowed (right-moving) linearly realized superconformal symmetry is $\CN = (0, 4)$ \cite{Nahm:1977tg, Hasiewicz:1989vp}. The failure of the R-symmetry to be chiral simply stems from the fact that  $p^i$ must transform under $Spin(8)$ in order that $Q^\da_m$ transforms as a spinor. This necessitates the inclusion of the nonchiral $L_{ij}$ piece in the R-symmetry generated by $J_{ij}$.

The action (\ref{type2action}) does admit a $\CN = (2, 2)$ and $\CN = (4, 4)$ superconformal symmetry if we are willing to forgo the $Spin(8)$ global symmetry. This fact is of particular physical significance in this context because it instructs us as to which of the symmetries of the near horizon of small black string we might hope to realize simultaneously and which not. Moreover, the $\CN = (2, 2)$ superconformal symmetry is the minimum that is required for us to be able to apply the Kraus-Larsen argument to obtain the Wald entropy correctly. We now like to exhibit this $\CN = (2, 2)$ superconformal symmetry and in particular, display the $12$ superconformal global symmetries for the right-movers of which we have only seen $8$ thus far, namely $Q^\da_0$.

For this purpose let us choose the embedding $SU(4) \times U(1)  \subset Spin(8)$ under which the spinor and the vector representations decompose as
\begin{equation}\label{decom}
    {\bf 8_s} = { \bf 4}^+ + { \bf \bar 4}^-, \quad
    {\bf 8_v} = { \bf \bar 4}^- + { \bf 4}^+
\end{equation}
where $\bf 4$ is the fundamental representation of $SU(4)$, $ \bf \bar 4$ its complex conjugate, and  superscript denotes the $U(1)$. One can now define the local $U(1)$ current $J$ as 
\begin{equation}\label{Juone}
    J_n =  (K^{12}_n + K^{34}_n + K^{56}_n + K^{78}_n) \, .
\end{equation}
and the supercurrents
\begin{equation}\label{supcurrents}
    G^{+}(\sigma^+) = \sqrt{\frac{1}{\a'}}\sum_a S^{a +} \partial X^-_{a}, \quad
     G^{-}(\sigma^+) = \sqrt{\frac{1}{\a'}} \sum_a S^{-}_{a} \partial X^{+a}
\end{equation}
where we have suppressed the worldsheet spin index and use the notation that $ S^{a +}$ transforms as ${ \bf 4}^+$ and $S^{-}_{a}$ as $ { \bf \bar 4}^-$ etc. It is easy to check that these modes of these currents along with $\CL_m$ and $J_n$ satisfy the usual $\CN =2$ superconformal algebra. In particular in addition to the Virasoro algebra (\ref{Vira}), we have
\begin{eqnarray}
  \{G^+_m, G^+_n\} &=& 0, \quad \{G^-_m, G^-_n\} = 0 \\
  \{G^+_m, G^-_n\} &=& 2 \CL_{m+n} + (m -n) J_{m+n} + \frac{c}{3}(m^2 -\frac{1}{4}) \delta_{m+n},\\
    \lbrack J_m, J_n \rbrack
      &=& {k} m\delta_{m+n}.
\end{eqnarray}
Note that the anomaly in the current current commutator  is proportional to $k = 2p$ which is related to the anomaly in the Virasoro algebra $c = 12p$. To see the global $OSp(2|2)$ algebra one has to use spectral flow and it is  easy to check that $(\CL_0 - J_0/2, \CL_{\pm 1}, J_0, G_0^{\pm}, G^-_{+1}, G^+_{-1})$ have the desired commutations. We see from here that $G_0^{\pm}$ generate the usual supersymmetries and $(G^-_{+1}, G^+_{-1})$ generate the conformal supersymmetries.

It is useful to summarize this construction of $\CN =2$ superconformal algebra using group theory.
We have chosen above  two linear combinations $G^{\pm}$ of the eight supercharges $Q^\da$, so that they form a closed algebra with the $J_n$.
The spinor  $Q^\da$ transforms under $\bf 8_c$.  When the $\bf 8_v$ and $\bf 8_s$ decompose as in (\ref{decom}), the conjugate spinor representation $\bf 8_c$ decomposes as
\begin{equation}\label{conj}
     {\bf 8_c} =  { \bf  1}^2  + {\bf 1}^{-2} + { \bf 6}^0
\end{equation}
We are discarding ${ \bf 6}^0$ and keeping only ${ \bf  1}^2$ and ${\bf 1}^{-2}$ in the form of $G^{\pm}$.
In the same way one can use the decomposition $SU(2)^4 \subset Spin(8)$ and use one of the $SU(2)$ as the conformal R-symmetry. Choosing an appropriate combination of the eight supercharges that transform as a doublet of this $SU(2)$ and discarding others, one obtains a closed $\CN=4$ superconformal algebra. It is not possible to construct a conformal R-symmetry that is larger than $SU(2)$ consistent with the fact that $\CN =4$ is the largest superconformal algebra that is allowed.

We can repeat this analysis for the other compactifications where the global $R$ symmetries are $Spin(d) \times Spin(8-d)$. In all the cases, the full global symmetry cannot be extended to linearly acting local currents and the maximal local $R$ currents possible is $SU(2)$ which corresponds to $\CN=(4,4)$ superconformal symmetry. 


\subsection{Holograms in the Internal Light Cone Gauge \label{Internal}}

We now discuss a somewhat unusual gauge that is a slight variant of the compact light cone gauge. The reason for considering this particular gauge will be clear after our bulk construction of boundary symmetries in the next section. We will encounter there commutation relations between the Virasoro symmetries and the rotational symmetries that are somewhat unusual but are quite natural from the point of view of the gauge that we call the `internal light cone gauge'.

Let us first consider the bosonic side of the heterotic string. This consists of bosons $\{ X^{M}, M = 0,..,9 \}$ and $\{ \tilde H^{I}, I = 1,..,16 \}$. Instead of picking the light cone gauge  to be (\ref{clc}), we define the bosons $Y, \bar Y = {1 \over \sqrt{2}} (X^{9} \pm \tilde H^{1})$.
We define the {\it internal light cone} coordinates to be
\begin{equation}\label{intlccoord}
    Y^{\pm} = \frac{( X^0 \pm Y)}{\sqrt{2}}.
\end{equation}
and fix the {\it internal light cone  gauge} to be
\begin{equation}\label{intlc}
    Y^{+} = y^+ + \frac{\alpha'}{\sqrt{2}}\left[(p^0+ \frac{1}{\sqrt{2}}( \frac{q}{R} + \frac{q'}{R'})   \tau  + \frac{1}{\sqrt{2} \alpha'}( p \, . R + p' \, . R' )\sigma  \right] \, ,
\end{equation}
where $R'$ is the radius of the internal boson and $(p', q')$ are the winding and momenta along this internal direction.
The remaining transverse fields are $\{X^{i}, (i =1...8), \bar Y, H^{m}, (m = 2...16) \}$ which we denote collectively by $ \phi^{a}, (a = 1...24)$.

As before, we can define dimensionless left-moving  lightcone momenta,
\begin{equation}\label{leftrightintLC}
    q^{+}_{L} = \sqrt{\frac{\a'}{2}}\left[ p^0 + \frac{1}{\sqrt{2}}( \frac{q}{R} + \frac{q'}{R'}) - {\frac{1}{\a' \sqrt{2}}( p \, . R + p' \, . R' }) \right]\, ,
\end{equation}
so that
\begin{equation}\label{xplusint}
  \partial_- Y^+ = \frac{\sqrt{\a'}}{2}q^+_L.
\end{equation}
We can solve as before the Virasoro constraints for the remaining longitudinal mode in terms of the transverse modes,
\begin{equation}\label{xminusint}
    \partial_- Y^- = \frac{\sqrt{\a'}} {q^+_L} ( T^{tr}_{--});
\end{equation}
where the superscript $tr$ refers to all spacetime and internal degrees of freedom that are transverse to $Y^{\pm}$.
In terms of modes, we have:
\be\label{phiminmode}
\T \a^{-}_{n} = {1 \over q^{+}_{R}} \left(  \half \sum_{a =1}^{24} \sum_{m=-\infty}^{\infty}  : \T \a^{a}_{n-m} \T \a^{a}_{m} :  -  \delta_{n} \right).
\ee
where $\T \a^{a}_{n}$ are the modes of the fields $\phi^{a}$.
$\T L_{n} \equiv q_{L}^{+} \T \a^{-}_{n}$ obey the Virasoro algebra with $c_{L}=24$.

So far, everything went like in the usual light cone gauge quantization. But since we did something funny, there are certainly differences.  Note that the $SO(32)$ or $E_{8} \times E_{8}$ gauge currents of the heterotic string involve the boson $H^{1}$. The manifest symmetries are therefore broken to $SO(28)$ and $E_{7} \times E_{8}$. The $SO(8)$ rotation symmetry on the other hand does not get enhanced -- the new transverse oscillator $\bar Y$ cannot be rotated into $X^{i}$ because of the non-zero winding around $X^{9}$.

Since we expect the theory to be independent of choice of gauge, we expect that the full gauge symmetry is actually restored.  The $SU(2) \subset E_{8}$ symmetry would be generated by the operators $\p_{-} \tilde H_{1}, e^{\pm i \sqrt{2} \tilde H_{1}}$ as before, except that these fields must now be expressed in terms of the appropriate  transverse modes. We will not study this in detail for now, and simply note the fact that
the currents involving $\tilde H^{1}$ will no longer be good conformal currents under the conformal algebra described above. For example, the  $U(1)$ current generated by
\bea\label{uonecurr}
\T j(\s - \tau) & \equiv & \p_{-} \tilde H^{1}
 =  \p_{-} { (Y - \bar Y ) \ \over \sqrt{2}} \\
& = &  \p_{-} \left( \half(Y^{+} - Y^{-})  -  {1 \over \sqrt{2}} \bar Y \right) \\
& = & {1 \over 4} p_{L}^{+} - \half \p_{-} \phi_{-}  -  {1 \over \sqrt{2}} \p_{-} \bar Y  .
\eea
The modes of this current
\be
\Rightarrow \T j_{n} =  {\delta_{n0} \over 4} p_{L}^{+} - \half \T \a^{-}_{n} -  {1 \over \sqrt{2}} \T \a^{\bar Y}_{n}
\ee
has a commutation relation with the conformal generators:
\be\label{uonecomm}
[\T \CL_{m} , \T j_{n} , ] =  \half m \T \CL_{n} +  {1 \over \sqrt{2}} n \T \a^{\bar Y}_{m+n}.
\ee
which is {\it not} the commutation relation of a usual  spin {\it one}
current which should have been
\begin{equation}\label{currentVir}
   [ \T \CL_{m}, \T j_{n}] = - n  \T j_{m+n}.
\end{equation}
This shows that even though the theory has manifest conformal affine symmetry in the compact light cone gauge, in this peculiar gauge,  the current is not
conformal. Thus, the conformal affine symmetry is not manifest and is broken by the gauge choice that mixes the compact direction with an internal one. 

Note that the zero mode of the current $\T j_{0} = \oint d \s \T j$ however continues to commute with the Hamiltonian and still remains as a manifest symmetry. If we wanted to quantize the string, this is all we need, since the only physical objects are integrated worldsheet currents. We are however looking for a macroscopic extended string where the local currents on the worldsheet are important.

For the superstring, we do a similar analysis. The internal direction in this case is a little more subtle and comes from within the fermionic lattice. We should then worry about how to construct the spin fields in order to get the spacetime supercharges and whether they transform correctly under the physical symmetries. We begin with the RNS fields $X^{M}, \psi^{M}$, $M = 0,1..9$.
We pick the fermions $\psi^{1,2,3}$ and bosonize two of them and get a system $(\theta, \psi^{\theta})$ where $\theta$  is at the free fermions radius. This breaks the symmetry from $Spin(10) \to Spin(7) \times Spin(3)$.

Defining as for the bosonic case the fields $Y, \bar Y = {1 \over \sqrt{2}} (X^{9} \pm \theta)$, we define the new internal light-cone coordinates and fix the internal light cone gauge as in (\ref{intlccoord}, \ref{intlc}). In addition, we also define the fermions  $(\psi^{Y}, \psi^{\bar Y})$ in an analogous manner and set
\be\label{intfermgge}
\psi^{+} =  {1 \over \sqrt{2}}  (\psi^{0} + \psi^{Y}) =0.
\ee
The remaining transverse fields are $\{X^{i},  (i =1...8),  \bar Y,  \psi^{m}, (m=1,..5), \psi^{\bar Y} \}$.

As for the bosonic case, we can solve the theory explicitly and find that the operators $\CL_{n} \equiv q_{R}^{+} \a^{-}_{n}$ obey  a Virasoro algebra with $c_{R}=12$. In addition, we can also solve for $\psi^{-} \equiv {1 \over \sqrt{2}}  (\psi^{0} - \psi^{Y})$ in terms of the other oscillators. As in the usual light cone gauge, the operators $G_{n } \equiv q^{+}_{R} \psi^{-}_{n}$ combine with the operators $\CL_{n}$ to form an $\CN=1$ superconformal algebra.

The manifest symmetries that remain in this  gauge are $Spin(5)$; the $Spin(3)$ which rotates the directions $(1,2,3)$ are broken by the choice of gauge.\footnote{There is still a rotation of the bosons $X^{1,2,3}$ but this is not the physical angular momentum that the string has in the usual light cone gauge which rotates both the bosons and fermions.}
As for the heterotic case, the $Spin(3)$ current which rotates the fermions which used to be a conformal current of weight one obeys an equation like (\ref{uonecomm}). The zero mode again is a manifest symmetry, and this can be added to the bosonic rotations to get back the Lorentz rotations.

To summarize, there is a global Lorentz rotation which is manifest in the internal light cone gauge. The local currents which rotates the chiral spinors on the worldsheet are however, not manifest symmetries.

Let us make a few comments about the Green-Schwarz spinors in this gauge.
We can consider the spinors which are formed by bosonizing the RNS fermions $\psi^{1,..8}$ and refermionizing them. As in $\S \, \ref{Internal}$, we can write the spinors as $S^{a\a}$ transforming as $(2,4)$ under the $Spin(3) \times Spin(5)$. The $SU(2)$ above which rotated $\psi^{1,2,3}$ can be rewritten as $S^{a\a} \s^{ij}_{ab} S^{b}_{\a}$. Our choice of gauge breaks this local symmetry on the worldsheet, but the zero mode is recovered as transforming well under the Virasoro algebra.

The first thing to note is that these spinors are {\it not} frozen by the choice of gauge. At the quantum level, we have set all the oscillator modes of the operator $\p_{+} ({1 \over \sqrt{2}} X^{0} + \half X^{9} + \half \theta)$ to be zero. However, the boson $\theta$ also enters the spinor lattice which is not fixed to be zero by this choice of gauge. The operators in the spinor lattice should of course be written in terms of the correct transverse modes.
The second observation is that under the Hamiltonian $L_{0}$,  the spinor currents are {\it not} dimension one half as one may have thought since the boson $\theta$ is not a free transverse oscillator.
Similar comments now apply for the supercharge $Q^{a \a}$ -- it seems to transform locally under the $Spin(3) \times Spin(5)$, but the $Spin(3)$ local chiral rotations are not manifest.
Because of the way the fermion is used in the choice of the light-cone, the construction of supercharges  is a bit subtle in this gauge and more work is needed to fully understand it. However, since it is just the familiar worldsheet theory in an unusual gauge, it is clear that such a construction must exist.

\section{Holographic Dual of the Type II Superstring \label{Bulk}}

We now consider the special case of $d=3$ for type II theory compactified on $\IR^{1,1} \times \IR^3 \times \bf T^{5}$ with the worldsheet of the macroscopic string  hologram extending along $\IR^{1,1}$.
Let us summarize the group theory associated with this compactification. The $Spin(8)$ symmetry  is broken by the toroidal identifications of the bosons to $Spin(3) \times Spin(5)$ where  $Spin(3)$ is generated by the angular momentum in the noncompact directions and  $Spin(5)$ acts on the tangent space of the torus.
The Green-Schwarz spinors transform as minimal spinors $(\bf 2, 4)$ under this decomposition, $S_{a \a}, a = 1,2, \a = 1,..4$, with a pseudo-Majorana condition  so that there are eight real degrees of freedom. Indices are raised and lowered with the matrices $\Omega^{ab}$ and $C^{\a \b}$, and the condition is $(S_{a \a})^{*} = S^{a \a} = \Omega^{ba} C^{\a \b} S_{b \b}$. The details are given in Appendix \ref{B}. The eight transverse bosons  are split as $X^{i}, X^{m}$ where $i=1,2,3$ and $m=4,..8$ are vector indices under $SO(3)$ and $SO(5)$.
The eight supercurrents $(Q_{a \a})_{s}$  obey the pseudo-Majorana condition $(Q_{a \a})^{*} = - Q^{a \a}$ with the opposite sign compared to the spinors $S^{a\a}$.

Since the rotational symmetry is broken only by some global identifications of the bosonic coordinates, the commutation relations for symmetry generators in the $d=3$ case  are closely related to the $d=8$ case discussed in $\S{\ref{Hologram}}$. One needs to simply decompose the $Spin(8)$ indices in terms of $Spin(3) \times Spin(5)$ indices as above, to rewrite the commutation relations  such as (\ref{GG}) or (\ref{JQ}) appropriately for the $d=3$ case.

From these commutation relations of the symmetries of the hologram, we expect the near horizon theory to have $Virasoro \times Virasoro$ symmetry. The hologram also instructs us that there should be a  $Spin(3)$ chiral symmetry current corresponding to the symmetry generated by the $K^{ij}_n$. We further expect the symmetries of $\bf T^5$ \footnote{We  use $\bf T^{5}$ to refer to a 5-torus as well its stringy symmetries depending on the context.} and at least eight supersymmetries that correspond to the zero modes $Q^{a \a}_0$. We will focus on only right-movers since the discussion is similar for the left-movers.

The Virasoro symmetry in the boundary is most naturally realized by having an $\bf AdS_{3}$ factor. String theory on $\bf AdS_{3}$ is by now well understood as a WZW model based on the $SL(2)$ current algebra on the worldsheet (see for example, \cite{Maldacena:2000hw, Maldacena:2000kv} and  references therein). We therefore start with $SL(2, \IR)$ super-affine algebra at level $k$ which factorizes into a bosonic $SL(2, \IR)$ affine algebra at level $k_b = k +2$ and three free fermions with total central charge
\begin{equation}\label{centralwzw}
    c = \frac{3 k_b}{k_b -2} + \frac{3}{2} = \frac{9}{2} + \frac{6}{k}.
\end{equation}
Given such a super-affine $SL(2, \IR)$ algebra in the bulk, there is an elegant construction due to Giveon, Kutasov, and Seiberg to obtain the boundary Virasoro algebra which has central charge
\begin{equation}\label{cb}
    c_R = 6 kp,
\end{equation}
where the integer $p$ naturally enters to be identified with the winding number. Since we want to identify it with the right-moving transverse superstring which has central charge $12p$, we are forced to the choice $k=2$ if we want agreement with the physical Wald entropy. The central charge of the $SL(2, \IR)$ factor for $k=2$ is thus $15/2$ from (\ref{centralwzw}). In addition, to account for the $\bf T^5$ factor, we must have five bosons and their NSR fermionic partners with total central charge $15/2$. Together these factors already account for all the central charge that is allowed for the rightmoving NSR superstring.
Our bulk worldsheet thus has a target space
\begin{equation}\label{target}
   SL(2, \IR)_{k=2} \times T^5.
\end{equation}
This particular Type-II model at level $k=2$ has been proposed earlier in the context of small black holes in \cite{Giveon:2006pr} who arrived at it from different considerations taking a limit of magnetically charged states. However, the  physical interpretation that we will advance here for the type II theory as well as for the heterotic theory in the next section will be substantially different especially with regards to the symmetries as we now summarize below.

Given a bulk worldsheet target space as in (\ref{target}), we are immediately led to a puzzle if we wish to identify it with the near horizon theory of a small black string. Since the allowed central charge of $c=15$ has already been used up, there is apparently no room for anything that can account for the rotational symmetries of the horizon.
It was even suggested in \cite{Giveon:2006pr} that these symmetries may completely disappear in the near horizon limit.

The existence of symmetries of the near horizon geometry of the fundamental string has been a confusing issue to understand at the level of supergravity solutions because the answer is hidden at the string scale.
Using the holograms as our guide proves to be very useful here. From the analysis of the boundary hologram for the $d=3$ case, there  is no doubt about the existence of $Spin(3) \times Spin(5)$ symmetry. As we have seen in $\S{\ref{Near}}$ , the existence of this symmetry is required also for understanding the R-symmetry and the  entropy through its relation to anomalies. Therefore, if we wish to identify the bulk theory (\ref{target}) as the holographic dual of the $d=3$ hologram, we  must correctly exhibit the symmetries of the hologram in particular the $Spin(3)$ generated by $K^{ij}_n$. Otherwise, we would be led to conclude that the holographic identification is incorrect.

It turns out that the rotational symmetries can be realized in a somewhat subtle  way using some special properties of the $k=2$ theory. For this purpose, we can view the target space theory as\footnote{This is a schematic but fairly standard notation, the product in this equation is not a direct product, there are constraints which tie the $U(1)$ to the coset.}
\begin{equation}\label{target3}
   \frac{SL(2, \IR)_{k=2}}{U(1)} \times U(1)  \times T^5
\end{equation}
This string background can be interpreted in a few different ways. In this paper, we always consider the time direction to be inside the coset, so that it is really a two dimensional black hole \cite{Witten:1991yr, Mandal:1991tz, Elitzur:1991cb}. Having said that, we note that all the calculations are done in an Euclidean setting with $H_{3}^{+}$ ({\it e.g.} \cite{Teschner:1997ft}) as standard in string theory, and one has to perform a Euclidean continuation.\footnote{As was emphasized in  \cite{Giveon:1998ns},  this is not such a big surprise -- the convergence of the path integral demands it just like in string theory in ten flat dimensions.} This spacetime has  zero temperature, and thus admits supersymmetry. 

Although the irrational nature of the above conformal field theory introduces many subtleties, in many  aspects, string perturbation theory can be understood as usual, in particular a modular invariant one-loop partition function can be written down. The partition function, symmetries and moduli space of precisely these theories for both the type II and heterotic cases were discussed in a slightly different context\footnote{The above target space admits a generalization in that the $U(1)$ factor can be made independent. This introduces more moduli at a particular value of which we recover the full $SL(2)$ structure. It also allows for a different Euclidean continuation where the time is outside the coset.} in \cite{Murthy:2003es, Murthy:2006eg}. 

Now precisely when $k=2$, the $U(1)$ boson happens to be at the free fermion radius so that the $U(1)$ symmetry is enhanced. The boson can then be fermionized into two fermions which together with its fermionic partner generates the $SU(2)_{k=2}$ current algebra. Using this current algebra on the worldsheet, one can then construct the symmetry currents. We discuss this construction of boundary Virasoro algebra and the boundary $Spin(3)$ symmetry in detail in $\S{\ref{Virasoro}}$ and $\S{\ref{Current}}$ using a particular (almost) free field representation.

The construction of symmetries raises a new puzzle. One finds that the commutation relations of the $SU(2)$ currents with the Virasoro generators are not what one would expect from the modes of a dimension one current. However, we find that the commutation are \textsl{precisely} as would be expected in an internal light cone gauge if the $SU(2)$ boson was used as one of the light cone direction.
This commutation leads us to identify the bulk theory defined by
(\ref{target}) as the holographic dual of the Type-II microscopic string hologram for $\bf T^5$ compactification but in the internal light cone gauge.

Another related issue that we clarify in this section is that of the construction of supersymmetries in the bulk theory. As we discussed in $\S{\ref{symholo}}$, the boundary hologram clearly has $(8,8)$ two dimensional supersymmetry. The supercharges we will construct commute with the Hamiltonian $\CL_{0}$
and are interpreted as the zero modes of the $(8,8)$ supercurrents in the R sector -- this implies \cite{Coussaert:1993jp} that the background is not pure global $AdS_3$ , rather the fermions in the bulk can have boundary conditions corresponding to a space which is not simply connected. An example of such a space is the extremal $J=M=0$ BTZ black hole \cite{Banados:1992wn, Banados:1992gq} which is singular in general relativity.  
The smooth string theory we have constructed seems to capture some aspects of the physics of this extremal black hole. It would be nice to understand the relation to earlier attempts to understand the entropy of this black hole using the symmetries of $AdS_{3}$ \cite{Strominger:1997eq,Carlip:1998qw}. 

A consistent interpretation of our theory\footnote{We thank Per Kraus for a clarifying discussion on this point.} is that it is the description of strings moving in the background which is one of the many Ramond ground states which make up the extremal massless BTZ black hole. As we shall see below, this vacuum carries maximal allowed R charge and can be identified with a smooth $AdS_3$ geometry with a constant gauge connection turned on which induces the fermions to change periodicity \cite{Lunin:2001fv,Maldacena:2000dr,Balasubramanian:2000rt}.

Let us make a quick comparison to the more familiar worldsheet construction of supersymmetric $\bf AdS_{3} \times S^{3}$ in \cite{Giveon:1998ns}. That construction gave rise to eight supercharges from the leftmovers (and another eight from the rightmovers) which formed among themselves the closed subalgebra involving the lowest $\left( \pm \half \right)$ modes of the supercharges of the $\CN=(4,4)$ superalgebra in the NS sector. Our construction of the spacetime supercharges is explicitly different.\footnote{Such a construction was written down in the appendix of \cite{Giveon:1998ns} where the discussion was restricted to theories where the $SU(2)$ currents and the $SL(2)$ currents do not mix on the worldsheet. The result was interpreted as a topological theory in spacetime after imposition of an additional constraint. In our case, an additional constraint is not required.} We  expand on this later when we discuss supersymmetry.

Having explained the issues involved, we  now present our discussion as follows. Starting with Type-II strings on (\ref{target}), we shall choose variables on the string worldsheet such that we can build the symmetries of $\bf AdS_{3} \times T^{5}$. In these variables, it will be clear that  there are also additional $SU(2) \times SU(2)$ symmetries in the system. These symmetries are ``stringy'' and replace the geometry at small scales.

\subsection{Superstrings on $\bf AdS_{3} \times T^{5}$}

Superstrings on $\bf AdS_{3}$ were studied in \cite{Giveon:1998ns} using the first order $\beta$-$\gamma$ system relevant to the $SL(2)$ symmetry. We instead use the fields $(\tau,\theta,\rho)$ mentioned above which are better suited to the symmetry of our problem  which near the boundary of the $\bf AdS$ space represent the global time, angular direction and the radial direction of global $\bf AdS_{3}$.\footnote{These variables were mentioned in \cite{Giveon:1998ns} and discussed in more detail in \cite{Hikida:2000ry, Giribet:2005xr}.} $(\tau,\theta)$ are free fields and $\rho$ has a linear dilaton of slope $Q=\sqrt{2 \over k}$ with central charge $c_{\rho}=1+3Q^{2}$. We  discuss a zero temperature supersymmetric $\bf AdS_{3}$ theory
which is Euclidean, correspondingly the $\tau$ direction will be compact. The symmetry algebra that we will obtain is the $SL(2)$ algebra with a timelike direction, and its infinite extension $Virasoro$. The correlation functions of the Lorentzian theory needs as usual, an analytic continuation.

Our variables must not be thought of as being the standard $\bf AdS_{3}$ variables\footnote{Note that our variables actually parameterize  a {\it flat} three dimensional solid cylinder in string frame.}; they are instead related to it by a ``T-duality'' discovered in \cite{Horowitz:1993jc} using Buscher's rules. The geometric action of this duality has a fixed point and actually even changes the boundary conditions -- but as we  discuss below, demonstrating the infinite dimensional Virasoro algebra associated with $\bf AdS_{3}$ elevates it to an exact stringy statement. These type of exact string backgrounds were introduced in \cite{Horowitz:1994ei, Horowitz:1994fv, Horowitz:1994rf}.

In addition, there are three fermions $\psi_{\tau},\psi_{\theta},\psi_{\rho}$ which make the worldsheet theory $\CN=1$ supersymmetric. We add the torus $\bf T^{5}$ represented by the free $\CN=1$ system $X^{i}, \chi^{i}, \, i=1..5$. All the directions are euclidean. $k$ is the supersymmetric level and the central charge $c= 3 + {6 \over k} + {3 \over 2} +  5 \times {3 \over 2}$.
To make this a critical string theory, we need add the $(b,c,\b,\gamma ) $ ghosts with $c=-15$. Demanding that the total central charge vanishes fixes $k=2$.

In these variables, there is a strong coupling singularity associated with the  $\rho$ direction. To keep string perturbation theory under control, we need to cap off this singularity. To do this, we notice that the variables $\rho, \tau, \psi_{\rho}, \psi_{\tau}$ have the central charge equal to that of the $\CN=2$ coset $SL(2)_{k=2}/U(1)$. This ``cigar'' coset has a geometry which smoothly caps off the strong coupling region, there is a modulus associated with the value of string coupling at the tip\footnote{In the actual $\bf AdS_{3}$ space, this modulus corresponds to the fixed value of the dilaton. }  which can be made small so that string perturbation theory is well-defined. This is summarized in Appendix \ref{A}. 

We have essentially spelt out the decomposition of the $SL(2)$ WZW model as $SL(2)/U(1) \times U(1)$ which has been used recently in many discussions of the $SL(2)$ model {\it e.g.} to understand spacetime supersymmetry \cite{Giveon:1999jg, Berenstein:1999gj}, to understand the spectrum  \cite{Pakman:2003cu}, the  partition function \cite{Israel:2003ry} and interactions \cite{Giribet:2001ft}.
Like other related representations, this one has its advantages and drawbacks. The manifest spacetime $SL(2)$ symmetry is lost, but we will recover explicit expressions asymptotically where we can use the above free field variables. In the full theory, we must use the coset algebra instead. On the other hand, the symmetries related to the other fields like $\theta, \psi_{\theta}$ is always manifest in our variables.

The symmetry algebra currents which we will write down are non-normalizable towards the weak coupling end as vertex operators on the worldsheet, and therefore act on worldsheet configurations which are localized in that asymptotic region. These correspond to inserting operators in the boundary in the AdS/CFT correspondence  \cite{deBoer:1998pp}. As for the $\beta$-$\gamma$ variables, this has precise implications for {\it e.g.} the understanding of the central charge \cite{Kutasov:1999xu}.

The angular direction of the cigar is at a specific radius which in our case is the free fermion radius $R=2$.\footnote{We  keep $\alpha'=2$ throughout this section.} Since this direction is associated with the Euclidean time direction, its compactness is not directly significant to us, however the Euclidean $\bf AdS_{3}$ geometry \cite{Giveon:1998ns} dictates that the angular direction $\theta$ also be at the same radius. This leads to an enhancement of symmetry which we discuss below.
For now, this implies that the vertex operators must have integer momenta in terms of this radius.

In the asymptotic region where the string coupling is small, the currents of the $\CN=1$ superconformal algebra are (with $Q=\sqrt{2 \over k}$):
\begin{eqnarray}\label{wsalg}
 T & = & -\half (\p \rho)^{2} - \half Q \partial^2\rho -\half (\p \theta)^{2} -\half (\p \tau)^{2} - \half (\p X^{i})^{2} \nonumber \\
& & \qquad \qquad - \half \psi_\rho \partial \psi_\rho   - \half \psi_\theta \partial \psi_\theta  - \half \psi_{\tau} \p \psi_{\tau} -\half \chi_{i} \p \chi_{i}  \\
 G & = & {\ i\over 2} \left(\psi_\rho\partial \rho + Q\partial \psi_\rho  +  \psi_{\theta} \p \theta +  \psi_{\tau} \p \tau + \chi_{i} \p X^{i} \right)  \nonumber
\end{eqnarray}
We choose exactly the same structure for the left movers.

\subsection{The $SL(2)_{R}$ symmetry from the worldsheet \label{Virasoro}}

If the above system indeed represents $\bf AdS_{3}$, it should be possible to find the infinite dimensional $SL_{2}$ symmetry algebra as operators built with these fields. Below, we construct such operators. As shown in \cite{Hikida:2000ry}, this is equivalent to the construction in  \cite{Giveon:1998ns}.\footnote{Our operators in (\ref{defJ}) and those in \cite{Hikida:2000ry} differ by a BRST exact operator which in the $(-1)$ picture is given by $e^{n(\tau + i \theta)} (\psi_{\tau} + i \psi_{\theta})$.}

Consider  the following dimension half operators on the worldsheet labeled by
$n \in \IZ$):
\be\label{defJ}
J^{(-1)}_{n} (z) \equiv e^{2n(\tau + i \theta)}(\half \psi_{\tau} - n \psi_{\rho}) (z)
\ee
which obey the following OPE's:
\bea\label{JOPEs}
G(z) J^{(-1)}_{n}(w) & \sim & {1 \over z-w} e^{2n(\tau + i \theta)} \left(\half \p \tau - n \p \rho + 2 n^{2} \psi_{\tau} \psi_{\rho} - i 2 n \psi_{\theta}(\half \psi_{\tau} - n \psi_{\rho}) \right) \equiv {1 \over z-w} J^{(0)}_{n} \nonumber \\
J^{(0)}_{n}(z) J^{(0)}_{m}(w) & \sim & -{nm+a \over (z-w)^{2}}e^{2n(\tau + i \theta)} (z) e^{2m(\tau + i \theta)} (w)  + {1 \over z-w} (n-m) J^{(0)}_{n+m}(w)
\eea
Firstly, note that the absence of higher order poles in the first OPE above involving the supercurrent  implies that all the currents $e^{-\varphi} J_{n}$ in the $(-1)$ picture ($J^{(0)}_{n}$ in the zero picture) are BRST invariant on the string worldsheet and thus act on physical string states. Secondly, the constant $a$ can be changed by adding the BRST trivial operator $e^{-\varphi}e^{n(\tau + i \theta)} (\psi_{\tau} + i \psi_{\theta})$ mentioned above. In the zero picture this is a total derivative, and its addition can be thought of as shifting the vacuum energy by a constant. This is also obvious from the Virasoro algebra written below -- the linear term in $n$ in the central extension can be reabsorbed in a constant shift of $L_{0}$. We  set this to unity as is the usual convention.

The charges $\CL_{n} =  \oint dz J^{(0)}_{n}$ obey the $SL_{2}$ spacetime algebra
\be\label{Vir}
[\CL_{m}, \CL_{n}] = (m-n)\CL_{m+n} + {c \over 12} (m^{3}-m) \delta_{m+n}.
\ee
The central term arises from the second order pole
\bea\label{centralterm}
\oint dw \oint dz {1 \over (z-w)^{2}} e^{2n(\tau + i \theta)} (z) e^{2m(\tau + i \theta)} (w) & = & \oint dw \left( \p_{w} e^{2n(\tau + i \theta)} (w) \right)e^{2m(\tau + i \theta)} (w) \nonumber \\
& = & n \oint dw \; 2 \p_{w} \left(\tau + i \theta  \right) e^{2(m+n)(\tau + i \theta)} (w) \nonumber \\
& = & n \delta_{m+n,0} \oint dw \; 2 \p_{w} \left(\tau + i \theta  \right) \nonumber \\
& \equiv & n \delta_{m+n,0} p .
\eea
We see that $c = 12 p$, where $p$ is measured by the integral (\ref{centralterm}) and is interpreted as the number of fundamental strings in the system \cite{Giveon:1998ns} . As explained in
\cite{Kutasov:1999xu}, this central charge computation done in a single string Hilbert space is the one measured by the long strings near the boundary of $\bf AdS$; the central charge measured by the short strings in the center of $\bf AdS$ arises from disconnected diagrams \cite{deBoer:1998pp}.

Note that if we want the $SL(2 \IR)$ currents above to be local with respect to the Hilbert space of states involving $(\tau , \theta)$, this needs the boson $\tau$ to be compact on a circle of the same size as the $\theta$ circle which we already have. The extrapolation from the semiclassical picture that we did above (\ref{wsalg}) thus seems to be consistent with the full quantum picture.
Recalling that the radius of the circle is tied intrinsically to the enhancement of symmetry, we can
restate the above statement as the following:  the consistency of the perturbative string theory with the correct symmetries produces exactly the expected entropy of the system. We take this as strong evidence for the existence of the hologram.

\subsection{The $SU(2)_{R}$ symmetry from the worldsheet \label{Current}}

As mentioned earlier,  we get an enhancement of symmetry since the boson $\theta$ is at the free fermion radius. The angular coordinate $\theta$ can be written in terms of two free fermions $e^{\pm i\theta} \equiv \irt2 (\psi^{1} \pm i\psi^{2})$, which along with the fermion $\psi^{\theta} \equiv \psi^{3}$ generates a left moving $SU(2)_2$ current algebra with currents $K^{i}(z)$ and corresponding charges $\CK^{i} = \oint K^{i}(z)$. This $SU(2)$ is a physical symmetry of the string theory as can be seen by the fact that its generators in the $(-1)$ picture given by the dimension half currents $\psi^{i}$ have a single pole with the worldsheet supercurrent (\ref{wsalg}).


From general arguments \cite{Kutasov:1999xu}, we expect that these symmetries would be extended to current algebras on $\bf AdS_{3}$ giving rise to an infinite set of conserved charges, just like the global $SL(2)$ is extended to the infinite dimensional Virasoro algebra  \cite{Brown:1986nw}.
However, it seems difficult to extend the $SU(2)$ global symmetry  in such a manner  -- the technique above of using the null operator $e^{n(\tau + i \theta)}$ naively fails because the boson $\theta$ which generates the $SU(2)$ zero modes is also involved in making the dimension zero operator $e^{n(\tau+ i \theta)}$.

One could try to define the infinite set of operators by using the OPE between the null operators and the $SU(2)$ currents above to define a normal ordering. This can be summarized in a nice way by defining the boundary currents $K^{i}(x)$ as an integral over the worldsheet weighted by a dimension zero operator $\Lambda(x,z)$ \cite{Kutasov:1999xu}. This is a nice exercise and $x$ gets the interpretation of parameterizing the worldsheet on the boundary.

But even if we do that, there seems to be a puzzle. One expects \cite{Kutasov:1999xu} that the worldsheet currents $K^{i}$ lead to corresponding currents in spacetime which are conformal currents under the spacetime Virasoro, {\it i.e.} the charges $\CK^{i}$ should be thought of as zero modes $\CK^{i}_{0}$ of a infinite set of charges $\CK^{i}_{n}$ which have the commutation relations
\be\label{cfmcurr}
[ \CL_{m}, \CK_{n} ] = - n \CK_{m+n}.
\ee
In particular, the zero mode should commute with all the Virasoro generators.
We can check that the expected commutation relation above (\ref{cfmcurr}) of a conformal current does not hold. For example, the commutation relations of the charges $\CK^{3}_{0} = i \oint \p \theta $ with the Virasoro charges (\ref{defJ}) is:
\be\label{noncfm}
[\CL_{m}, \CK^{3}_{0}]  = m \CL_{m}
\ee
This puzzle is resolved by noting that this commutation relation is precisely the one in the internal light cone gauge of the boundary theory in the previous section!

To summarize, on the bulk string worldsheet, the $Virasoro$ generators and the $SU(2)$ generators mix, this makes the conformal nature of the $SU(2)$ currents in spacetime non-manifest.
The holographic dual of this statement is that the choice of the internal light cone gauge on the boundary string breaks the conformal nature of the $SU(2)$ current in the same way.
It would be very interesting to understand if there is a different formulation of the theory where the choice of gauge is not built in, but can be added and the change of gauge is covariant.

The identification of the $SU(2)_R$ symmetries above also allows us to identify the spacetime vacuum more precisely. The integral (\ref{centralterm}) tells us that the spacetime vacuum carries the maximal allowed $U(1)_R \subset SU(2)_R$ charge and therefore should be interpreted as the unique vacuum in the Ramond sector with the corresponding value of the charge. 

\subsection{The $\bf T^{5}$ symmetries from the worldsheet}

The translation symmetries associated to the $\bf T^{5}$ at a generic point in its moduli space can be also be extended into a level one $U(1)^{5} \times U(1)^{5}$ current  algebra in spacetime. The right and left moving operators for these symmetries are:
\be\label{torus}
\CP_{n}^{iR} =  \oint dz \, e^{-\varphi} \chi^{i} e^{n(\tau + i \theta)} (z) ,
\ee
and a similar one for left movers.

\subsection{The supersymmetries from the worldsheet}

Now that we have understood the bosonic symmetries fully from the worldsheet point of view, we turn to the supersymmetries. The supercharges must live in representations of the bosonic symmetries discussed above.

To get spacetime supersymmetric theories, the standard procedure in the case of compactifications to flat space is to use an $\CN=2$ algebra on the worldsheet \cite{Friedan:1985ge}.  In the case of theories  on $\bf AdS_{3}$, it was pointed out \cite{Giveon:1998ns} that  the algebra expected from the boundary superconformal theory is actually reproduced in the bulk using a {\it different} construction wherein one simply makes spin fields out of ten free fermions and keeps those that are physical and mutually local.\footnote{For the case $k=2$, there {\it does} exist a different $\CN=2$ structure which reproduces these supercharges as we briefly mention below. This is not true for generic $k$ \cite{Israel:2003ry}.}

We  actually use the standard procedure of \cite{Friedan:1985ge} using the $\CN=2$ worldsheet structure.\footnote{Such a construction was sketched in the appendix of \cite{Giveon:1998ns}, and was interpreted (after an additional projection which threw out four of the eight supercharges) as a possible description of the R sector of the $\CN=(4,4)$ algebra of the D1/D5 system on $\bf T^{4}$. As we have discussed, the bosonic as well as the super symmetries in our boundary theory are explicitly different.}
This ensures that the supercharges we build are physical operators. The spacetime supercharges we thus obtain are indeed {\it not} those of the NS sector of a boundary $\CN=(4,4)$ algebra, but instead the supercharges of a $\CN=(8,8)$ superalgebra which have zero conformal dimension. This is in accord with the discussion of the hologram in $\S{\ref{Hologram}}$.

To proceed, we spilt the worldsheet fields into two groups. The first consists of the cylinder formed by $\rho , \theta, \psi^{\rho}, \psi^{\theta}$. This has an $\CN=2$ algebra (\ref{ntwodil}) with a $U(1)_{R}$ symmetry $J^{1}_{R} \equiv i \p \phi = - i \psi^{\rho} \psi^{\theta} + i \p \theta$. This is summarized in Appendix \ref{A} with $X \equiv \theta$.\footnote{Note that in the above construction, since $\tau$ and $\theta$ are at the same free fermion radius, there is a different $\CN=2$ supersymmetry on the worldsheet where $X\equiv \tau$ is fermionized, and $\psi^{\theta}$ is paired with $\psi^{5}$ from the torus.
Using this structure to build the supercharges gives the standard construction of \cite{Giveon:1998ns} for the case $k=2$ wherein the eight supercharges have conformal dimension $\pm \half$ and form part of a spacetime $\CN=(4,4)$ algebra. Note that the two sets of eight supercharges are not local with respect to each other, so we have to choose one or the other.}

The rest of the fields $\tau, X^{i}$ and their superpartners $\psi^{\tau}, \chi^{i}$ $(i=1..5)$ are paired up to get a complex structure and a corresponding $\CN=2$ structure. The fermions can be bosonized $\p H_{1} \equiv \psi^{\tau} \chi^{1}$, $\p H_{2} \equiv \chi^{2} \chi^{3}$, $\p H_{3} \equiv \chi^{4} \chi^{5}$.
The $U(1)$ $R$ current is then expressed as a sum of the bilinears in these fermions $J^{2}_{R} \equiv i( \p H_{1} + \p H_{2} + \p H_{3})$.

To perform a chiral $\IZ_2$ projection, we can use the symmetry generated by the $U(1)_{R}$ current $J^{1}_{R} + J^{2}_{R}$. In practice, the GSO projection is best implemented by introducing target space supercharges and demanding locality of physical operators, as in \cite{Kutasov:1990ua}.
We introduce the $(1,0)$ supercurrent operator
\be\label{defsusy}
S_{A}(z)=e^{-{\varphi \over 2} \pm i{\phi\over 2}} S_{a}(z)
\ee
where $S_{a}$ is the spin field of $SO(6)$ built out of the three pairs of free fermions and $\varphi$ is the bosonized superghost.  There are $2^{4}=16$ such supercurrent operators, and $8$ of them are mutually local, these are all of one chirality in the six dimensions. Of course, we would have obtained the same supercharges by simply making spin fields out of our ten free fermions and demanding consistency. For the type II theories, there is also a similar condition on the leftmoving side giving rise to the IIA or IIB theories.

Now, we don't really have a $SO(6)$ symmetry and we must arrange our supercurrents in the $SU(2) \times Spin(5)$ symmetry. From the reduction of $Spin(6) \to Spin(5)$, it is clear that the supercurrents are spinors under $Spin(5)$. One can also check that they are spinors under the $SU(2)$ -- recall that the $K^{3}$ of the $SU(2)$ is given by $K^{3} = \oint dz \, \p \theta$, the supercharges above all have a $\theta$ dependence in the exponent with a coefficient $\pm \half$.
We then have eight mutually local supercurrents which fall into the minimal spinor of this group which is a $(2,4)$ with a (pseudo)reality condition using the antisymmetric charge conjugation matrices $\Omega_{ab}$ and $C_{\a \b}$ as described in $\S \, \ref{Internal} $ and Appendix \ref{B}. We  accordingly call the supercurrents $S^{a \a} (z)$.

Note that the supercurrents are local on the worldsheet with respect to all the vertex operators generating the spacetime bosonic symmetry currents  described earlier, in particular the spacetime Virasoro currents -- again, we note the special nature of the $k=2$ theory, this does not happen for generic $k$ as was discussed in  \cite{Berenstein:1999gj}.

The algebra of the supercharges $Q_{a \a} = \oint dz \, S_{a \a}(z)$ can be deduced by examining the OPE of the currents (\ref{defsusy}) above. After performing the usual picture changing operation on the right hand side, we get:
\be\label{susyalg}
\{Q_{a \a}, Q_{b \b} \}=  2 \Omega_{ab} C_{\a\b} L_{0} + 2  \Omega_{ab} (C\gamma^i)_{\alpha\beta} P_i.
\ee
where $(i = 1..5)$, and $L_{0} = \oint \p \tau$ .
Since there is no $\tau$ dependence of the supercurrents (\ref{defsusy}), it is clear that all the supercharges have vanishing conformal dimension.
If we restrict to the subspace where $P^{i}=0;\; i=1..5$, we get  the supersymmetry algebra discussed earlier.

These supercharges are dimension zero under the spacetime Virasoro algebra (\ref{Vir}), but they involve the boson $\theta$, and hence suffer from the same problem as the $SU(2)_{R}$ symmetry -- the supercurrents in spacetime seem {\it not} to be dimension half conformal currents. Again, this is what is seen in the boundary theory in the internal light cone gauge.

\section{Holographic Dual of the Heterotic String \label{Heterotic}}
The heterotic string shares with the type II string a chiral set of fields and physics governed by these fields are similar. In this section, we shall try to emphasize the novel features of the heterotic theory arising from the leftmovers and the process of combining the two chiralities of the string fluctuations. 
From the leftmovers, we expect chiral symmetry currents of $E_{8} \times E_{8} \times Virasoro$. We then have the same bosonic fields $\rho, \tau, X^{m}; m =1, \ldots, 5 $, as on the right with $c=10$ and the gauge lattice of $E_{8} \times E_{8}$ or $SO(32)$ with $c=16$\footnote{For brevity we will often refer only to $E_{8} \times E_{8}$ but our considerations apply to both possibilities.}. This gives us already a total central charge of $26$. Counting the central charge as before, this means that the target space must be of the form\footnote{This conformal field theory looks similar to the one used in \cite{Giddings:1993wn}. However, as we shall see below, the string theory is different. In particular, the theory we consider is supersymmetric and has no background monopole charge.}
\begin{equation}\label{target4}
        \frac{SL(2, \IR)_{k_b =4}}{U(1)}  \times T^5 \times E_8 \times E_8,
\end{equation}
To build a heterotic string theory, we need to combine these left movers with the rightmovers of (\ref{target3}) with $k_{b}=k+2$. For generic values of $k$, these heterotic cosets have not been studied very well, but it is known that the radius of the left moving boson generating the $U(1)$ is related to the radius of its right moving counterpart by a factor of $\sqrt{k_{b}/k}$ which is $\sqrt{2}$ in our case \cite{Giveon:2006pr}. In the case of $k=2$, we actually can understand this better -- the modular invariance of the partition function dictates that the left moving boson $\T \theta$ generating the $U(1)$ must be at the self dual radius  (consistent with the above factor of $\sqrt{2}$) so that the symmetry is enhanced to $SU(2)_{1}$  \cite{Murthy:2006eg}.

If we wish to now use a construction similar to in the previous section to construct the Virasoro symmetry in the boundary, we would require such a boson $\tilde \theta$ at self dual radius. Since we would like the torus to have free moduli corresponding to its radii, it must actually arise from within the $E_{8} \times E_{8}$ lattice  in the same way that it arose from the $SU(2)_{2}$ represented by three free fermions  for the supersymmetric side. This will  non-trivial consequences which seem strange at first sight, but as we shall see, simply corresponds to a corresponding  gauge choice in the boundary theory as in  $\S{\ref{Internal}}$.

\subsection{The $SL(2)_{L}$ symmetry from the worldsheet}

In the heterotic theory, the form of the generators on the bosonic side are different -- they are actually much simpler since there is no constraint arising from $\CN=1$ worldsheet supergravity.
The form of the $SL_{2}$ currents is very similar to the supersymmetric case (\ref{defJ}, \ref{JOPEs}), but is simpler. For a boson $\T \theta$ with canonical normalization $\T \theta(z) \T \theta(0) \sim - \ln{\zb}$, using the techniques of the previous section, it can be checked that the currents
\be\label{defJhet}
J_{n}(\zb) \equiv  e^{{n \over a} (\T \tau + i \T \theta)} \pb (a \T \tau -n \T \rho)  (\zb)
\ee
are physical and obey the Virasoro algebra with central charge $c=6 \kappa p$ with $\kappa =  {2 a^{2}}$, $p = {1 \over  a} \oint \p (\T \tau + i \T \theta)$.


As discussed above, the value of $a = \sqrt{2}$,   i.e. all the allowed operators are of the form $\CV e^{i {n \over \rt2 }\T \theta}$. This implies that for the heterotic side, $\kappa = 4$ and $c=24 p$.
The central charges are then simply $c= 6 \kappa p$; with $\kappa =2$ and $\kappa=4$ for the supersymmetric and heterotic sides. As was noted in  \cite{Giveon:2006pr}, this is consistent with the fact that the level of the supersymmetric coset and the bosonic coset are $k=2$ and $k_{B}=k+2$ for the two theories.
The interpretation of this fact in   \cite{Giveon:2006pr} was in terms of a ``thermodynamic''  entropy wherein  the cigar angular variable is the Euclidean time in a finite temperature theory. 
It is not very clear what such an interpretation means when the radius of the circle\footnote{More precisely, the operator content of a boson at the given radius.} on the left and the right are not equal like in the heterotic case above. 

The microscopic computation of the entropy above follows from the central charge computation in the Virasoro algebras on the left and the right.  The asymmetric nature of this circle direction is completely consistent with the factorization of the theory into left and right movers. The theory is at zero temperature   consistent with supersymmetry; there are two non-interacting Hamiltonians and the two corresponding central charges $c_{L,R}$ arise simply from counting the various vacuum configurations. It is indeed interesting that for the type II case, such a thermodynamic calculation of the entropy agrees with our microscopic one. It would be nice to understand this better.

As mentioned in the previous section, our interpretation of the resulting spacetime is also different. There is a  stringy  $Spin(3)$ symmetry and maximal supersymmetry as discussed in detail there.

\subsection{The $E_8 \times E_8$ symmetry from the worldsheet}

The same thing can be repeated for the gauge generators to get an $E_{8} \times E_{8}$ ($SO(32)$)  spacetime current algebra
\be\label{currents}
\CJ^{ab}_{n}  =  \oint d\zb \, J^{ab} (\zb) e^{{n \over \rt2}(\T {\tau} + i \T {\theta})} (\zb)
\ee
where the $J^{ab}(\zb)$ are the dimension one gauge currents on the worldsheet.

Note that because of the way we have `borrowed' $\tilde \theta$ from the gauge lattice only  the $E_{8} \times E_{7}$ (or $SO(28)$) part of the gauge currents are realized as conformal affine algebra. The $SU(2) \subset E_{8} \times E_{8}$ generated by the boson $\T \theta$ suffers from the same problem as the $SU(2)_{R}$, i.e. it seem to be non-conformal in spacetime. This we interpret as for the $SU(2)_R$ symmetries to be a consequence of a particular gauge that we have chosen for this construction which seems to correspond to an internal light cone gauge. The `non-conformal' commutators between the $SU(2)$ currents (\ref{currents}) and Virasoro generators are then exactly what one expects in this particular gauge. In fact, we could have embedded the $SU(2)$ in the original gauge lattice in many different ways which should probably be interpreted as different possible gauge choices.
Note however that the global $E_8 \times E_8$ symmetry generated by the zero modes of the currents commutes with the Hamiltonian and hence one can surely assert the existence of the full  $E_8 \times E_8$ symmetry.

\section{Conclusions and Open Problems \label{Conclusions}}

We have a proposed that a simple, free two dimensional SCFT living on a macroscopic superstring can be regarded as the hologram for the gravitational theory on $\bf AdS_{3}$ in the vicinity of a macroscopic string.
For the $\bf T^5$ compactification, we have written down the holographic dual as an exact worldsheet in the bulk. As we have seen the logic of this construction is very tight which we summarize below.
\begin{enumerate}
  \item To realize the full Virasoro symmetry of the boundary  theory it is natural to have at least an  $SL(2, \IR)$ WZW model. A bulk construction of a Brown-Henneaux conformal algebra of the $\bf AdS_{3}$ can then be given incorporating the winding number $p$. This Virasoro algebra has the correct central charge expected from the Wald entropy only if the level of the WZW model is $k=2$.
  \item For the heterotic string, the consistency  of the perturbative string theory demands that the level of the left moving WZW model be $k_{b}=k+2 = 4$, which gives the correct left moving central charge.
  \item If in addition we have a $\bf T^5$ factor for the superstring and an additional level one $E_8 \times E_8$ factor for the left-movers of the heterotic string, then one finds that the maximally allowed worldsheet central charge of $15$ for the superstring and $26$ for the bosonic string are already saturated. The target space therefore must be of the form (\ref{target3}) for the superstring and (\ref{target4}) for the left-moving heterotic string.
    \item The form of the target space in (\ref{target3}) however raises an important puzzle about the symmetries. If this is to be identified with a small black string in a $\IR^3 \times T^5$ compactification then its  global symmetries must contain a  $Spin(3)$  factor corresponding to $\IR^3$ rotations. Fortunately, precisely for $k=2$, the $U(1)$ boson in $(\ref{target3})$ is at the free fermion radius and it is then possible to construct the $Spin(3)$ currents using this fact.      All global symmetries expected for the horizon of a black hole and independently from the boundary hologram can be constructed from the bulk theory.  Similarly, for the heterotic string, the boson is at the self dual radius which makes it possible to recover all the symmetries.
  \item One also expects  $Spin(3)$ affine currents from the bulk corresponding to $K^{ij}_n$ in the hologram. The symmetry currents can be constructed from the bulk but one finds that the commutation relations with the Virasoro generators are unusual and are not what one might expect for the modes of a conformal dimension one current. We note however that the commutations  are \textit{exactly} what one might expect from the boundary hologram (\ref{uonecomm}) if it was gauge fixed using an unusual internal light cone gauge discussed in $\S{\ref{Hologram}}$.
     This suggests that one should identify the symmetry algebra constructed from the  bulk constructed using these particular variables with the corresponding algebra in the hologram in a particular internal light cone gauge.
  \item One can construct in the bulk eight chiral supersymmetries corresponding to the zero modes $Q^{a \a}_0$ expected from the boundary in the Ramond sector.      
   \item To obtain a small black hole from a small black string, we should identify along the length of the string to obtain a compact circle. The generator of such a translation is $\CL_0 - \tilde \CL_0$. Note that both $\CL_0$ and $\tilde \CL_0$ commute with the $Spin(3)$ and $E_8 \times E_8$ currents and hence such an identification would commute with the symmetries.
\end{enumerate}
It is nontrivial that a  such a consistent  worldsheet theory exists.
The bulk worldsheet construction is very tightly constrained. The requirements of  the maximal allowed central charge of the bulk worldsheet and the physical requirements following from symmetries and Wald entropy lead almost uniquely to the theory that we have used.
Using this theory we are then able to give a detailed construction of all boundary symmetries in a particular free field realization which seems to correspond to a particular choice of internal gauge in the boundary theory. 

The most unsatisfactory part of our construction is the necessity to choose this particular unusual gauge. If the basic identification of the target space (\ref{target3}) and (\ref{target4}) is correct, then it should be possible to construct the symmetries in a way that corresponds to the usual (compact) light cone gauge where they are manifest.  This suggests that it may be possible to generalize the GKS construction \cite{Giveon:1998ns} to construct the boundary Virasoro algebra abstractly only from the $SL(2, \IR)/U(1)$ factor that does not require us to borrow a $U(1)$ factor. The coset theory does  not admit $SL(2, \IR)$ symmetries. However, what we are really after are is a Virasoro symmetry in the boundary. The coset theory is expected to have an extended chiral algebra. For example, in the compact analog, $SU(2)_k/U(1)$ is just the parafermion theory that does not have $SU(2)$ symmetry but admits a conserved spin-3 currents that generates the $W_3$ algebra which is nonlinear. Perhaps one can obtain  realization of the boundary Virasoro algebra utilizing these additional (nonlinear) symmetries in the bulk.  This is a very interesting open problem and could be related to large extended algebras as suggested in \cite{StromingerStrings} (see \cite{Kraus:2007vu} for a recent discussion).

The existence of a worldsheet construction resolves many of the puzzles relating to small black holes and in particular gives a construction of the near horizon geometry of both heterotic and Type-II small black holes in four dimensions. The identification of the macroscopic string worldsheet theory as a boundary hologram is very useful in understanding the physics. In particular, the issues about global and local symmetries and the applicability of the Kraus-Larsen argument in this context becomes transparent. There are chiral stringy currents generated by $K^{ij}$, a linear combination of some of which can be identified with an R-current. These do not correspond to the nonchiral gauge symmetries generated by $J^{ij}$ that are visible in supergravity for which the bulk Chern-Simons terms vanishes.

There are a number of possible generalizations and open questions.
\begin{enumerate}
  \item The holograms  make it clear that there is nothing special about $d=3$. This is consistent with what one might expect from scaling analysis in supergravity. So if a holographic dual exists for $d=3$, it is expected to exist for all values. It seems likely that the other higher dimensional theories can be obtained simply by decompactifying the $\bf T^5$. This is what is required in the boundary hologram and it should be true also in the bulk. For example, when $\bf T^5$ is replaced by a noncompact $\IR^5$, one can add to the angular momentum $J^{mn}$ an orbital angular momentum term involving $L^{mn}$. The full $Spin(8)$ symmetry is not manifest but it is because of the choice of the gauge that breaks it to $Spin(8)$ to $Spin(3) \times Spin(3)$. Getting the off-diagonal currents $J^{mi}$ should also be possible but requires more work.

  \item One can also contemplate holograms in more general compactifications. This gives a rich class of examples of this kind of `string-string holography' where the hologram  and the dual both have a worldsheet description. For example, $\bf T^5$ could be replaced by $\bf K3 \times S^1$.  The analysis of the hologram can be repeated simply by replacing $\bf T^4$ by $\bf K3$ in the transverse CFT of the macroscopic superstring. The holographic dual is obtained by a similar replacement in (\ref{target3})
      so that the target space is now
      \begin{equation}\label{target5}
      \frac{SL(2, \IR)_{k=2}}{U(1)} \times U(1)  \times S^1 \times K3.
      \end{equation}
      Since both the bulk and the boundary are given by tractable worldsheet conformal field theories, it ought to be possible to test this holography in greater detail than has been possible in other contexts. For example, a comparison of correlation functions might be possible as was done for the related F1-NS5 system \cite{Gaberdiel:2007vu, {Dabholkar:2007ey}}.
 
  \item  One  thing to keep in mind is that the boundary theory is expected to correspond to  a string field theory of the bulk theory that includes the multi-string states as well. 
      It would be interesting to see if it is possible to construct the string field theory of the bulk using conventional methods of string field theory and to compare it with the boundary hologram. In another related direction, it would be interesting to try to understand the known non-perturbative objects like D-branes in the bulk $AdS_{3}$ from the perspective of the boundary theory.

 \item  In the bulk, the $\bf AdS_{3}$ structure in the bulk is an important part of the symmetry algebra, which manifests itself in the associated Brown-Henneaux stress tensor. In the boundary, this translates to closed algebras which contain the  Virasoro generators. As we have seen, with a few additional assumptions, like that of linear realization and no higher spin currents constrains these algebras very tightly. It would be interesting to look for non-linear generalizations which involve higher spin operators. 

 \item There is also the related issue of single string {\it v/s} multi string Hilbert spaces. The boundary theory has a large extended chiral algebra which involves all the chiral operators on the string, these are not expected to be realized in the bulk single string Hilbert space. The symmetries of the single-string Hilbert space in the bulk  form a closed subalgebra. So it seems reasonable to expect that only a maximal closed algebra that includes  Virasoro will be realized in the single-string Hilbert space and not all extended algebras.
      
 \item  The orbital angular momentum generators $L^{ij}$ which rotate only the bosons on the boundary appears to be absent in the bulk. On the boundary, for the heterotic case, there are two  symmetries generated by $J^{ij}$ and $S^{ij}$ but in the bulk we have only one. More work is needed to fully understand the details of this correspondence. 
 
  \item  A singly macroscopic string with $p$ windings along a single circle is marginally unstable under decay into $p$ strings with unit winding. One then has to take into account the multi-string branch analogous the Coulomb branch in the D1-D5 system \cite{Seiberg:1999xz}. In the context of F1-NS5 systems this necessitates turning on Fayet-Illiapoulos terms in the gauge theory that correspond to RR fluxes in the bulk.
      For fundamental strings, the multi-string branch can be prevented by the simple device of adding momentum or winding along in internal, but not both, along an internal circle. This makes the configuration  stable under marginal decay without changing the entropy as explained in $\S{\ref{Near}}$. For such configurations, an internal light-cone gauge would be more natural.
\item The way we measured entropy is by using a long fundamental string probe in $\bf AdS_{3}$. This involved a definition (although natural)  of the ``number'' of F-strings $p$ which in the context of $SL(2, \IR)$ current algebra is related to the spectrally flowed sectors. 
 
   \item The hologram that we have discussed  can also be related to the usual gauge-gravity duality \cite{Maldacena:1997re, Gubser:1998bc, Witten:1998qj} by S-duality. If we consider $N$ D1-branes in this context \cite{Itzhaki:1998dd}, then the dilaton becomes strong near the core. So one must perform  an S-duality trasnformation to go to the weak coupling  F-string description to see the horizon that we have discussed. We are taking a deep infrared limit of the D1-brane worldvolume theory. In this limit the $1+1$ theory is simply the symmetric product  $(\IR^8)^N/S_N$ where $S_N$ is the symmetric group of $N$ objects \cite{Dijkgraaf:1997vv}. There are many twisted sectors of this orbifold classified by the conjugacy classes of the symmetric group which are given by collections of cycles of various lengths (see \cite{Hikida:2000ry, Hosomichi:1998be, Argurio:2000tb} for a discussion  in a similar context). Here we have discussed the sector in the orbifold with cycle length $p$. 
\end{enumerate}

\subsection*{Acknowledgements}

It is a pleasure to thank Edi Gava, Gaston Giribet, Per Kraus, Finn Larsen, Samir Mathur, Shiraz Minwalla, Kumar S. Narain, Seif Ranjbar-Daemi, Andy Strominger, Ashoke Sen, Sandip Trivedi, Walter Troost for valuable discussions. We also thank Andy Strominger and Sandip Trivedi for early collaboration.

\appendix

\section{A short summary of the $\CN=2$ $SL(2)/U(1)$ coset \label{A}}
The semiclassical geometry of the cigar in the string frame is:
\bea\label{cigmetric}
ds^2 &=& d\rho^2+\tanh^2({Q\rho\over 2}) dX^2, \qquad X \sim X +{4\pi\over Q}; \\
\Phi &=& \Phi_{0} - \log \cosh({Q\rho\over 2}), \qquad B_{ab}=0.
\eea
with the string coupling $g_s=e^\Phi$. Semiclassically, this metric is a good one for string propagation because the dilaton obeys the equation $2D_aD_b\Phi + R_{ab}=0$, where $D_a$ is the spacetime covariant derivative, and $R_{ab}$ is the spacetime curvature. After adding  two fermions, the theory can be made $\CN=2$ supersymmetric. The exact description is given by the supersymmetric coset $SL(2)_{k}/U(1)$ with $Q^{2} = 2/k$. The string coupling at the tip $g_{s} = e^{\Phi_{0}}$ is the one modulus of this theory. To get the two dimensional Lorentzian black hole, we need to Wick rotate the angular direction $X = i T$.

In the asymptotic region $\rho \rightarrow \infty$, the geometry reduces to a flat cylinder of radius $R={2\over Q}$ with the dilaton varying linearly along its length.
The currents of the $\CN=2$ superconformal algebra in this region are
 \bea\label{ntwodil}
 T &=&-\half (\partial \rho)^2 -\half (\partial X)^2 - \half
 (\psi_\rho\partial \psi_\rho + \psi_X \partial \psi_X)
 - \half Q \partial^2\rho \nonumber \\
 G^\pm &=& {i\over 2} (\psi_\rho \pm i\psi_X)\partial(\rho \mp
 iX) +{i\over 2} Q\partial (\psi_\rho \pm i\psi_X) \\
 J &=& -i\psi_\rho\psi_X +iQ\partial X \equiv i\partial H
 +iQ\partial X \equiv i\partial \phi \nonumber
 \eea
In the main text, we use this $\CN=2$ structure  to build spacetime supercharges.

In the theory on the cylinder, a generic rightmoving operator looks like $\CO = \CV e^{i k X + (p - {Q \over 2})\rho}$, where $\CV$ depends on the rest of the coordinates. The dimension of this operator is $\Delta_{\CV}  + {k^{2} \over 2} + {Q^{2} \over 8} - {p^{2} \over 2}$.

\section{Spinor Conventions \label{B}}

We follow the conventions of \cite{VanProeyen:1999ni}.
$Spin(3) \times Spin(5)$ spinors  have two indices $\lambda_{a \a}$ with $a$ in the ${\bf 2}$ of $Spin(3)$ and $\a$ in the ${\bf 4}$ of $Spin(5)$. In five Euclidean dimensions, one can pick the charge conjugation matrix $C$ to be $\s_{1} \otimes \s_{2} $ which obeys $C^{t} = -C$, $C^{\dagger} = C^{-1} = C$, $C \G C^{-1} = + \G^{t}$. The sigma matrices are the standard ones: $\s_{1} = \pmatrix{0 & 1 \cr 1 & 0 }$ ,  $\s_{2} = \pmatrix{0 & -i \cr i & 0 \\}$, $\s_{3} = \pmatrix{1 & 0 \cr 0 & -1 \\}$, and the convention for direct product is such that $C = \pmatrix{0 & \s_{2} \cr \s_{2} & 0 }$.

The matrix $B=C^{t}$ is used to impose a Majorana type condition. In this case, we have a pseudo-Majorana condition using the matrices $B_{\a \b}$ and $\Omega_{ab} = i \sigma_{2}$. The pseudo-Majorana condition is $\lambda^{*} = B \Omega \lambda$, such that $(B^{*})^{*} = B$ Another way to write this is to define the Majorana conjugate as $\bar \lambda = \lambda^{t} \Omega^{t} C$ and then define $\bar \lambda = \lambda^{\dagger}$.

Spinors have a lower index $\lambda_{\a}$ and their Majorana conjugates have an upper index $\lambda^{\a}$. Indices are raised and lowered with $C$ and $\Omega$:
\bea\label{spinorind}
\lambda^{a \a} = \lambda^{a}_{\b} C^{\b \a} \; ; &\quad & \lambda_{\a}^{a} = C_{\b \a} \lambda^{a \b} \\
\lambda^{a \a} = \lambda^{\a}_{b} \Omega^{a b} \; ; &\quad & \lambda_{a}^{\a} = \Omega_{a b} \lambda^{\a b}
\eea
In terms of the indices, the pseudo-reality condition is $(\lambda_{a \a})^{*} = \lambda^{a \a}$.

\bibliographystyle{JHEP}

\bibliography{hologram}

\end{document}